\documentclass[12pt,preprint]{aastex}
\usepackage{multirow}
\usepackage{booktabs}
\usepackage{graphicx}
\usepackage{amsmath}
\usepackage{longtable}
\usepackage{rotating}
\usepackage{enumerate}

\newcommand{\msun}{M_\odot}
\newcommand{\mj}{M_{\rm J}}
\newcommand{\ms}{M_{\rm S}}
\newcommand{\mn}{M_{\rm N}}

\newcommand{\rsun}{R_\odot}
\newcommand{\rp}{r_{\rm p}}
\newcommand{\mplanet}{M_{\rm p}}
\newcommand{\mt}{M_{\rm thermal}}

\begin{document}
\title{How Bright are Planet-Induced Spiral Arms in Scattered Light?}

\shorttitle{Spiral Arms in NIR Images}

\shortauthors{Dong \& Fung}

\author{Ruobing Dong\altaffilmark{1} \& Jeffrey Fung\altaffilmark{2}}
\affil{Department of Astronomy, University of California, Berkeley}
\altaffiltext{1}{NASA Hubble Fellow, rdong2013@berkeley.edu}
\altaffiltext{2}{NASA Sagan Fellow, jeffrey.fung@berkeley.edu}

\clearpage

\begin{abstract}

Recently, high angular resolution imaging instruments such as SPHERE and GPI have discovered many spiral-arm-like features in near-infrared scattered light images of protoplanetary disks. Theory and simulations have suggested that these arms are most likely excited by planets forming in the disks; however, a quantitative relation between the arm-to-disk brightness contrast and planet mass is still missing. Using 3D hydrodynamics and radiative transfer simulations, we examine the morphology and contrast of planet-induced arms in disks. We find a power-law relation for the face-on arm contrast ($\delta_{\rm max}$) as a function of planet mass ($\mplanet$) and disk aspect ratio ($h/r$): $\delta_{\rm max}\approx((\mplanet/\mj)/(h/r)^{1.38})^{0.22}$. With current observational capability, at a 30 AU separation, the minimum planet mass for driving detectable arms in a disk around a 1 Myr 1$\msun$ star at 140 pc at low inclinations is around Saturn mass. For planets more massive than Neptune masses, they typically drive multiple arms. Therefore in observed disks with spirals, it is unlikely that each spiral arm originates from a different planet. We also find only massive perturbers with at least multi-Jupiter masses are capable of driving bright arms with $\delta_{\rm max}\gtrsim2$ as found in SAO 206462, MWC 758, and LkH$\alpha$~330, and these arms do not follow linear wave propagation theory. Additionally, we find the morphology and contrast of the primary and secondary arms are largely unaffected by a modest level of viscosity with $\alpha\lesssim0.01$. Finally, the contrast of the arms in the SAO 206462 disk suggests that the perturber SAO 206462 b at $\sim100$ AU is about $5-10\mj$ in mass.\end{abstract}

\keywords{protoplanetary disks --- planets and satellites: formation --- circumstellar matter --- planet-disk interactions --- stars: variables: T Tauri, Herbig Ae/Be --- stars: pre-main sequence}


\section{Introduction}\label{sec:intro}

In recent years, scattered light imaging at near-infrared (NIR) wavelengths has revealed intriguing spiral-arm-like features in many protoplanetary disks, such as AB Aur \citep{hashimoto11}, MWC~758 \citep{grady13, benisty15}, SAO 206462 \citep{muto12, garufi13, stolker16}, HD 142527 \citep{avenhaus14, canovas13}, HD 100453 \citep{wagner15hd100453}, HD 100546 \citep{boccaletti13, avenhaus14hd100546, currie14, currie15, garufi16}, AK Sco \citep{janson16}, LkH$\alpha$ 330 \citep{akiyama16}, and V1247 Orionis \citep{ohta16}. The leading hypothesis for the origin of spiral arms is that they are excited by planets embedded in disks. Alternative explanations, such as gravitational instability\citep[e.g.,][]{pohl15, dong15giarms, kratter16} and radially bend shadows caused with finite photon travel time \citep{kama16}, have been investigated, but may not explain the observed arms so far.

Making quantitative predictions on the properties of the arm-driving planets is crucial to the study of planet formation. Because these planets are still located in their formation environment, their masses, orbits, and ages provide immediate constraints on planet formation models. The first attempt in this direction was made for SAO 206462 by \citet{muto12}, in which the linear theory of planet-induced density waves \citep[Equation 44 in][]{rafikov02} was applied to fit the shape of the two observed arms to derive possible locations of two arm-driving planets. This linear theory fitting scheme was subsequently adopted in many other studies \citep[e.g.,][]{grady13, benisty15, vandermarel16sao206462, janson16}. However, explaining observed spiral arms with linear wave theory is questionable due to the following three problems. 
\begin{enumerate}
\item In the linear theory, one planet only excites one continuous spiral density wave (an inner part + an outer part, e.g., Fig. 4 in \citealt{ogilvie02}). In several cases a pair of nearly symmetric arms are present (e.g., MWC 758, SAO 206462, LkH$\alpha$ 330), demanding two planets at symmetric locations, which triggers dynamical stability issues.
\item In the linear theory, once waves are exited by planets at Lindblad resonance locations, they propagate as sound waves. Therefore, their openness (i.e., pitch angles) increases with an increasing ratio of sound speed $c_{\rm s}$ to the local Keplerian speed, or equivalently disk aspect ratio $h/r$. Many observed spiral arms are wide open, demanding unreasonably high disk temperature in order to fit them with the linear theory \citep[e.g., MWC758, ][requiring $h/r>0.2$]{benisty15}.
\item The linear wave theory only applies to low mass companion -- planets with masses $\mplanet$ much lower than the disk thermal mass $\mt$,
\begin{equation}
\mplanet\ll \mt=M_\star\left(\frac{h}{r}\right)^3,
\label{eq:mthermal}
\end{equation}
where $M_\star$ is the mass of the central star. Once $\mplanet$ approaches $\mt$, nonlinear effects kick in and the linear theory breaks down \citep{goodman01, rafikov02}. In the linear wave regime, the amplitude of the wave is required to be much lower than unity,
\begin{equation}
\frac{|\Sigma-\Sigma_0|}{\Sigma_0}\ll 1;
\label{eq:mthermal}
\end{equation}
where $\Sigma$ and $\Sigma_0$ are perturbed and unperturbed (initial) surface density, respectively. The detectability of such weak waves was put into question by \citet{juhasz15}.
\end{enumerate}

Recently, numerical simulations of disks with a single massive companion ($\mplanet\gtrsim\mt$, $\sim$Jupiter mass or above) have shown that multiple arms can be excited by one such companion \citep{juhasz15, dong15spiralarms, zhu15densitywaves, fung15, dong16armviewing, bae16sao206462}. The excitation and propagation of these arms are fully nonlinear, and they are well detectable in NIR imaging, with morphologies closely resembling the observed ones. The proof of concept of this giant-companion-multiple-arm scenario was provided by HD 100453, in which the double-arm-driving companion is detected \citep{wagner15hd100453, dong16hd100453}. Additionally, these arms have been shown to have significantly larger pitch angles at a given $h/r$ than the linear theory prediction, resolving the above Problem \#2.

The contrast of planet-induced density waves in surface density has been well studied both (semi-)analytically and numerically. \citet{goodman01} showed that in the linear regime ($\mplanet\ll\mt$),
\begin{equation}
\frac{\Sigma_{\rm peak}-\Sigma_0}{\Sigma_0}\propto \frac{\mplanet}{\mt} \left(\frac{x}{h}\right)^{0.5},
\label{eq:sigma_contrast}
\end{equation}
where $\Sigma_{\rm peak}$ is the peak surface density on the wave at each radius, and $x$ is the radial distance from the planet's orbit. This result has been verified using sharing box simulations by \citet{dong11linear, dong11nonlinear, zhu13}, and using global simulations by \citet{duffell12}. In the nonlinear regime ($\mplanet\gtrsim\mt$), \citet{zhu15densitywaves} have shown that the contrast of the wave ($(\Sigma_{\rm peak}-\Sigma_0)/(\Sigma_0)$) reaches order unity.

Quantitative connections between the arm morphology to the planet mass, a key quantity to planet formation theory but very hard to measure, was first made by \citet{fung15}, who linked the symmetry of arms excited by one planet to the planet's mass. What remains unknown is how the arm contrast (or strength) in scattered light, a direct observable, depends on the mass of an arm-driving planet. In general, the visibility of waves in scattered light is determined by the location and curvature of the disk surface at the waves' locations \citep[e.g.,][]{takami13}, which are subsequently determined by both the wave amplitude in surface density, and the distribution of material in the vertical direction. The latter is determined by balancing gravity, gas pressure, and vertical kinematic support.

In this paper, we carry out 3D global numerical hydrodynamics and radiative transfer simulations (Section~\ref{sec:setup}) to study the contrast of the arms as functions of planet mass and disk scale height (Section~\ref{sec:results}); based on which, we explore what kind of planets can drive visible arms, and comment on existing observed spiral arm systems (Section~\ref{sec:discussions}). In particular, we ask one specific question: is it possible to explain the observed arms with low mass planets each responsible for one arm as describable by the linear theory (Section~\ref{sec:minimum})? This general topic has been previously touched by \citet{rosotti16} and \citet{juhasz15}, who found planets of tens of Earth mass are not able to drive detectable density waves. However, in both works the hydro calculations were done in 2D. \citet{zhu15densitywaves} have shown that the vertical kinematics of the waves can significantly enhance their visibility in scattered light, thus 3D hydro simulations are needed. We further extend the \citet{zhu15densitywaves} results to low mass planet regime (Section~\ref{sec:2d3d}). Finally, we briefly discuss the prospects of spiral arm imaging with next generation telescopes in Section~\ref{sec:tmt}, before we summarize the main take aways in Section~\ref{sec:summary}.


\section{Simulation Setup}\label{sec:setup}

\subsection{Hydrodynamics Simulations}\label{sec:hydro}

We use the 3D Lagrangian-remap hydro code \texttt{PEnGUIn} \citep{fung15thesis} to perform our hydro simulations in spherical coordinates. We denote $r$, $\phi$, and $\theta$ as the usual radial, azimuthal, and polar coordinates. The cylindrical coordinates $R$ and $z$ are then simply $R=r\sin\theta$ and $z=r\cos\theta$. \texttt{PEnGUIn} solves the following compressible Euler equations:
\begin{align}
\label{eqn:cont_eqn}
\frac{{\rm D}\rho}{{\rm D}t} &= -\rho\left(\nabla\cdot\mathbf{v}\right) \,,\\
\label{eqn:moment_eqn}
\frac{{\rm D}\mathbf{v}}{{\rm D}t} &= -\frac{1}{\rho}\nabla p  - \nabla \Phi \,,
\end{align}
where ${{\rm D}}/{{\rm D}t}$ is the Lagrangian derivative, $\rho$ is the gas density, $\mathbf{v}$ the velocity field, $p$ the gas pressure, and $\Phi$ the combined gravitational potential of the star and the planet. We adopt a locally isothermal equation of state, such that $p= c_{\rm s}^2 \rho$, where $c_{\rm s}$ is the local sound speed of the gas. In the frame of the star, and for a planet on a fixed, circular orbit, $\Phi$ is:
\begin{equation}\label{eqn:potential}
\Phi = -G\left[\frac{M_*}{r}-\frac{\mplanet}{\sqrt{r^2 + r_{\rm s}^2 + \rp^2 - 2R\rp\cos{\phi'}}} - \frac{\mplanet R\cos{\phi'}}{\rp^2}\right] \,,
\end{equation}
where $G$ is the gravitational constant, $M_*$ the stellar mass, $r_{\rm s}$ the smoothing length of the planet's potential, and $\phi' = \phi-\phi_{\rm p}$ denotes the azimuthal separation from the companion. The third term in the bracket is the indirect potential due to the acceleration of the frame. We fix $M_*=\msun$ and vary $\mplanet$. We choose $r_{\rm s}$ to be $10\%$ of the planet's Hill radius: $r_{\rm s} = 0.1\rp (\mplanet/3 M_*)^{1/3}$. The planet is introduced into the simulation gradually over the first orbit. We carry out both inviscid and viscous simulations; in the latter the viscosity is characterized by the \citet{shakura73} $\alpha$ parameter (inviscid simulations have $\alpha=0$).

\subsubsection{Initial and boundary conditions}\label{sec:initial}
The initial density profile reads:
\begin{align}
\label{eqn:rho_vertical}
\rho =& \rho_{\rm mid} \exp\left(\frac{GM}{c_{\rm s}^2} \left[\frac{1}{r}-\frac{1}{R}\right]\right) \,, \\
\label{eqn:rho_radial}
\rho_{\rm mid} =& \rho_0 \left(\frac{R}{\rp}\right)^{-\frac{9}{4}} \,,
\end{align}
where $\rho_{\rm mid}$ is the midplane density, and $\rho_0 = 1$.\footnote{This value is re-normalized in the radiative transfer simulation (Section \ref{sec:mcrt}).} The sound speed profile reads:
\begin{equation}\label{eqn:initial_c}
c_{\rm s} = c_{\rm s,0} \left(\frac{R}{\rp}\right)^{-\frac{1}{4}} \,,
\end{equation}
where $c_{\rm s,0}$ is the sound speed at the planet's radial position. The disk aspect ratio $h/r$ can be expressed as $c_{\rm s}/v_{\rm k}$, where $v_{\rm k}$ is the Keplerian speed. This choice of sound speed profile produces a flaring disk, since $h/r\propto R^{1/4}$ is increasing with $R$. Combining with Equation \ref{eqn:rho_radial}, we have a surface density profile that goes as: $\Sigma\sim\rho_{\rm mid}h\propto R^{-1}$. Because $h/r$ determines the thermal mass $\mt$ of the disk, we perform two sets of simulations using $h/r=0.05$ ($\mt=0.125\mj$) and $h/r=0.1$ ($\mt=1\mj$), evaluated at the planet's location, to investigate its effects. Radiative transfer calculations show that in an externally irradiated disk around a typical Herbig Ae/Be star with a stellar mass 2.5~$M_\odot$, a stellar radius  2~$R_\odot$, and a photosphere temperature $10,000$~K, the midplane temperature $T_{\rm mid}$ is about 65~K at 30~AU, corresponding to $h/r\sim0.06$; while in a disk around a typical T Tauri star with 0.5~$M_\odot$, 2~$R_\odot$, and $4000$~K, $T_{\rm mid}$ is about 30~K and $h/r\sim0.08$.

The initial velocity field assumes hydrostatic equilibrium, and has zero radial and polar velocities. The orbital frequency $\Omega$ is therefore:
\begin{equation}\label{eqn:omega}
\Omega = \sqrt{\Omega_{\rm k}^2 + \frac{1}{r}\frac{\partial p}{\partial r}} \,,
\end{equation}
where $\Omega_{\rm k}$ is the Keplerian frequency.

Our simulation domain spans 0.3 to 2.0 $\rp$ in the radial direction, the full $2\pi$ in azimuth, and from $0^{\circ}$ to $19^{\circ}$ ($35^{\circ}$) in the polar direction when $h/r = 0.05~(0.1)$, where $0^{\circ}$ is the disk midplane. This top polar boundary corresponds to 7 scale heights above the midplane at the planet's location, and is about twice as high as the disk scattering surface in our radiative transfer simulations (Section~\ref{sec:mcrt}). The dimensions of our simulation grid is $320(r)\times1024(\phi)\times64(\theta)$, where the radial cell sizes are logarithmic, and the azimuthal and polar cell sizes are uniform. Both the inner and outer radial boundaries are fixed at their initial values. For the polar boundaries, we enforce symmetry at the midplane since we only simulate the upper half of the disk, and use a reflecting condition at the top to prevent mass from entering or leaving the simulation domain. 

It takes only a few orbits for the spirals waves to be fully established, since the sound crossing time is only about 3 (1.5) orbits when $h/r=0.05~(0.1)$. With similar setups, in \citet{dong16armviewing} we simulated the spiral arms up to 100 orbits, and found the arm morphology and contrast change little between 10 and 100 orbits. In this paper we simulate all models up to 30 orbits. Our main results (e.g., the correlations in Figure~\ref{fig:contrast_raw}) do not change between 10 orbits and 30 orbits. To minimize the effects of gap opening and emphasize the arms, we present results at 10 orbits. For massive planets, gaps reach a substantial depth after tens of orbits but can only be fully opened over a much longer period than the timescale considered here. This may weaken the contrast of the arms, as part of the arms closer to the planet may be truncated, which only strengthens our conclusions about the minimum detectable planet masses (see below).

Figure~\ref{fig:sigma_contrast} shows the surface density perturbation in the arms at $r=0.7\rp=21$~AU for both $h/r$ series, where $\Sigma_{\rm peak}$ is the peak surface density and $\Sigma_{\rm background}$ is the azimuthally averaged surface density. As discussed in Section~\ref{sec:intro} (Eqn~\ref{eq:sigma_contrast}), for $\mplanet\lesssim\mt$, density perturbation ($\Sigma_{\rm peak}/\Sigma_{\rm background}-1$) is roughly proportional to planet mass, while for $\mplanet\gtrsim\mt$ density perturbation saturate at order unit.

\subsection{Radiative Transfer Simulations}\label{sec:mcrt}

The 3D hydro density grid is subsequently read into the \citet{whitney13} 3D Monte Carlo radiative transfer (MCRT) code to produce synthetic polarized intensity (PI) images at $H$-band. This procedure is the same as in \citet{dong16armviewing}. For the central source we assume a $T=4350$K and $R=2.325\rsun$ star, appropriate for a $1\msun$ star at 1 Myr \citep{baraffe98}. We scale the hydro grid so that the planet is at $\rp=30$ AU (thus the outer disk boundary is at 60 AU). Given the inner working angle in the latest scattered light observations by SPHERE and GPI ($\sim0\arcsec.1$ or 14 AU at 140 pc, the distance to Taurus), this is about the closest that a planet can be and still make the inner arms observable. We note that where the planet is does not affect the relative contrast of the planet induced features in full resolution images; its radial location only matters when convolving the images under a given angular resolution. In the radial direction $r$, the MCRT grid starts from the dust sublimation radius $r_{\rm sub}$, where the disk temperature reaches 1600 K. The 3D MCRT grid is identical to the hydro grid, except in the inner region inside the hydro inner boundary, where we use the same hydro polar and azimuthal gridding, and logarithmic gridding in the radial direction with 50 grid points. 

3D Hydro simulations produce gas volume density, while scattered light is determined by the distribution of small dust, generally sub-micron in size. Such dust is dynamically well coupled with the gas, therefore we linearly scale the gas density $\rho_{\rm gas}$ to obtain dust density $\rho_{\rm dust}$. Between $r_{\rm sub}$ and the hydro inner boundary (9 AU), we assume an axisymmetric inner disk smoothly joining the outer hydro disk with $\Sigma\propto r$. We assume interstellar medium (ISM) dust \citep{kim94} for these NIR scattering small dust, which have a power law size distribution $n(s)\propto s^{-3.5}$ between $s=0.002-0.25~\micron$. Grain growth in disks can convert smaller dust into bigger particles. If the maximum dust size grows to 1 mm, our assumed small dust population would account for $\sim$2\% of the total dust mass. Their scattering properties are calculated using the Mie theory. The total dust disk mass within 60 AU is scaled to $10^{-5}\msun$.\footnote{This corresponds to a total disk mass of $0.001\msun$ if the gas-to-dust-mass ratio is 100:1 and all the dust are in these small sub-$\micron$-sized dust. If dust have grown to a maximum size of 1 mm while maintaining the same power law size distribution and gas-to-dust-mass ratio, the corresponding total gas disk mass would be $\sim0.06\msun$.} We note that the specific ways of filling up the inner disk (as long as it is axisymmetric and not casting a shadow), the detailed assumptions on the properties of the small grains, as well as the assumed dust mass $\pm$ one order of magnitude do not affect the relative contrast of the arms at face-on viewing angle (dust scattering angle is $\sim90^\circ$ everywhere).

MCRT simulations produce raw images at ``full resolution'', only limited by the finite pixel size. To compare with real data, we convolve these images by a Gaussian point spread function (PSF) to achieve an angular resolution of $0\arcsec.04$ assuming the objects are at 140 pc, comparable with what SPHERE and GPI are doing at NIR wavelengths. All images are produced using 4 billion photon packets, and all will be shown in linear stretch. 


\section{Results}\label{sec:results}

We simulate in total 12 models with planet masses varied between 1 Neptune mass and 3 Jupiter masses, and $h/r$ at the planet's location of either $0.05$ or $0.1$ (Table~\ref{tab:models}). Ten models are inviscid and two are viscous. To refer to these models, we denote $\mn$ (``MN'') as Neptune mass, $\ms$ (``MS'') as Saturn mass, $\mj$ (``MJ'') as Jupiter mass, and ``H5'' and ``H10'' as the $h/r=0.05$ and $0.1$ models respectively. For the two viscous simulations, we use ``A2" and ``A3" to indicate $\alpha=10^{-2}$ and $10^{-3}$, respectively. Figure~\ref{fig:basic3} shows a basic set of simulation results for the 1MJ-H10 model. The planet is located at [$x=0,y=30$]~AU in all models (except Section~\ref{sec:sao}). The central $0\arcsec.07$ is blocked to mimic the effect of an inner working angle. All results in this section are at face-on view angle. We reserve the gray color scheme for surface density maps and red color scheme for polarized $H$-band images in all cases.

\subsection{2D Hydro Simulations Produce Artificially Weak Inner Arms in Scattered Light}\label{sec:2d3d}

\citet[][Figures 18 \& 19]{zhu15densitywaves} showed that the non-linear spiral shocks excited by a massive companion ($\mplanet\gtrsim \mt$) in 3D hydro simulations are significantly different in scattered light images when compared to spirals shocks in 2D simulations with an assumed vertical hydrostatic density profile (i.e. puffing up a 2D disk). Here we show that the same is true in the linear / weakly-nonlinear regime ($\mplanet<\mt$) as well.

Figure~\ref{fig:2d3d} compares the image synthesized directly from a true 3D hydro simulation, and the image synthesized from puffing up a 2D hydro surface density map, for the 1MS-H10 model ($\mplanet=0.3\mt$). Clearly, the inner spiral arms are more prominent in the former. This is because 3D hydro simulations produce significantly different density structures in the vertical direction in the inner waves comparing with the vertical hydrostatic solution achieved in the background disk (Equation \ref{eqn:rho_vertical}), due to the vertical kinematic support in the waves. This is consistent with previous 3D hydro simulations showing significant vertical motion in planet-perturbed disks \citep[e.g.,][]{fung16, bae16swi}. As shown in the bottom panel, at the inner arms 3D hydro density decreases slower with increasing distance from the midplane; in other words, material is ``pushed'' up to raise the disk surface; therefore the waves receive more starlight, and become more prominent. On the other hand, the outer spiral arms are less prominent in the 3D than in the 2D simulation, suggesting a ``suppression" of vertical density structure. In the 3D simulation the outer arms are very weak, making them unlikely to be detectable. More detailed analysis of the kinematic support is deferred to a forthcoming paper.

Combining with the results in \citet{zhu15densitywaves}, the presence of vertical kinematic support necessitates 3D hydro simulations in studies of observational signatures of planet-driven spiral arms for all planet masses, including both the linear and non-linear regimes.

\subsection{One Arm, Two Arms, Three Arms, Many Arms}\label{sec:arms}

Multiple inner arms driven by a single companion have been robustly seen in the regime of $\mplanet\gtrsim\mt$. In this section we show that they can also be present in the weakly nonlinear regime where $0.1\mt\lesssim\mplanet\lesssim\mt$. We focus on the inner arms as they are more likely to be detectable, and emphasize that the outer arms may or may not have the same number of arms as the inner ones.

Figure~\ref{fig:3mn1mj3mj} shows the full resolution image of three H10 models: $3\mn$ ($0.16\mt$), $1\mj$ ($1\mt$), and $3\mj$ ($3\mt$). Strikingly, even at $0.16\mt$, two inner arms with roughly equal strengths are clearly visible in the $H$-band image (although the secondary is less prominent in the surface density). At the lowest thermal mass case (1MN-H10, or $0.05\mt$), we do not find robust evidence for the secondary and additional arms. A fully consistent theory to explain the generation of addition spiral arms is still under development \citep{lee16}. 

A tertiary arm excited by a massive companion has been seen by both \citet{zhu15densitywaves} and \citet{fung15}. Here it is clearly present in the $\mplanet=\mt$ and $\mplanet=3\mt$ models. While both the primary and secondary arms in images become brighter with increasing $\mplanet$, the strength of the tertiary varies. At $\mplanet=\mt$, although it is only marginal in the surface density map, the tertiary achieves about the same strength as the primary and secondary in scattered light; at $\mplanet=3\mt$, it is more recognizable in the surface density, but much weaker than the primary and secondary in scattered light. This variation is likely caused by varying kinematic support, poorly understood at the moment. Additional features, which may be considered as a $4^{\rm th}$ (and more) arm, can also be found in the $1\mt$ and $3\mt$ models.

\subsection{Arm Contrast in Scattered Light}\label{sec:contrast}

In this section, we study the contrast of the arms as functions of location, planet mass, and disk aspect ratio. We note that because we do not include viscosity, the wave amplitudes in our simulations are not affected by dissipation (except from shocks). In addition, our synthetic images have no noise (except the intrinsic Poisson noise, which is negligible), therefore spiral arms in our full resolution images have the maximum possible contrast.

\subsubsection{The Azimuthal Arm Profiles}\label{sec:azimuthal}

Figure~\ref{fig:azimuthal_1mj_h10} shows the azimuthal surface brightness (SB) profiles for both the full resolution and convolved 1MJ-H10 images at $r=0\arcsec.11$ to $0\arcsec.19$, and Figure~\ref{fig:azimuthal_conv} shows the azimuthal profiles for different $\mplanet$ for convolved images at $r=0\arcsec.15$. In both images of 1MJ-H10, 3 major arms can be clearly traced out on the profiles as 3 distinct peaks (the full resolution image has larger variations for obvious reasons); their position angles (PA) shift with distance, and their relative strengths vary with distance. As $\mplanet$ increases, the azimuthal separation between the primary and secondary arm increases, consistent with the results of \citet{fung15}. Meanwhile the relative strengths of these arms change with $\mplanet$ (i.e., the tertiary arm marked by the third peak on the profile becomes weaker at large $\mplanet$), as discussed in Section~\ref{sec:arms}.

\subsubsection{The Radial Arm Profiles}\label{sec:radial}

Figure~\ref{fig:radial_max} shows the arm contrast at each radius, defined as the ratio of the maximum over azimuthally averaged surface brightness,
\begin{equation}
\delta(r) = \frac{\rm Max\ SB({\it r})}{\rm Azimuthally\ Averaged\ SB({\it r})}
\label{eq:contrast_r}
\end{equation}
for all convolved images (each azimuthal profile in Figures~\ref{fig:azimuthal_1mj_h10} and \ref{fig:azimuthal_conv} contributes one data point at the given $r$). We find that $\delta(r)$ increases with $\mplanet$, as more massive planets drive stronger waves. The peaks of $\delta(r)$ at $r<\rp$ and $r>\rp$, denoted as $\delta_{\rm max,inner}$ and $\delta_{\rm max,outer}$, correspond to the maximum contrasts of the inner and outer arms, respectively (marked for the 3$\mj$ $h/r=0.1$ model). In nearly all models, $\delta_{\rm max,inner}>\delta_{\rm max,outer}$, except in the 1MJ-H5 ($\mplanet=8\mt$) and 3MJ-H5 ($\mplanet=24\mt$) models, two simulations with the most massive planet in unit of thermal mass. Considering that the outer disk is intrinsically fainter in absolute SB due to its larger distance from the star, and that a massive companion may truncate the disk to further suppress the outer arms, the inner arms are more likely to be identified. This echoes one of the conclusions of \citet{dong15spiralarms}. For this reason, we focus on the inner arm contrast in the next section. Qualitatively the Figure~\ref{fig:radial_max} results hold for full resolution images as well.

\subsubsection{Global Arm Contrast as a Function of Planet Mass}\label{sec:contrsat-mp}

Figure~\ref{fig:contrast_raw} shows the global maximum of $\delta(r)$ for the inner arms, denoted $\delta_{\rm max}$, as a function of $\mplanet$ for full resolution images. Because waves are excited at Lindblad resonances close to the planet and damp through shock dissipation at large distances, the peak contrast is expected to occur at some finite distance away from the planet's orbit. Simulations show $\delta_{\rm max}$ occurs at around 4 scale heights from the planet's orbit ($r\sim0.8\rp=24$ AU $=0\arcsec.17$ for the $h/r=0.05$ series, and $r\sim0.6\rp=18$ AU $=0\arcsec.13$ for the $h/r=0.1$ series). In general, for a given $h/r$, $\delta_{\rm max}$ increases with increasing $\mplanet$, as more massive planets excite stronger waves; for a given $\mplanet$, $\delta_{\rm max}$ decreases with increasing $h/r$, as waves are stronger in thinner disks. Figure~\ref{fig:contrast_sigma_image} shows $\delta_{\rm max}$ as a function of $\Sigma_{\rm peak}/\Sigma_{\rm background}$ at the corresponding location for each model. Overall, higher contrast in surface density is closely traced by higher contrast in scattered light.

With both $h/r$ profiles, $\delta_{\rm max}$ appears to be in a tight power law correlation with $\mplanet$ with a power index $\sim$0.2. A formal fitting produces (the dashed lines):
\begin{align}
\label{eq:contrast_h5}
\delta_{\rm max} = 2.5 \left(\frac{\mplanet}{\mj}\right)^{0.23},\ {\rm for\ } h/r=0.05,\\
\label{eq:contrast_h10}
\delta_{\rm max} = 2.0 \left(\frac{\mplanet}{\mj}\right)^{0.22},\ {\rm for\ } h/r=0.1.
\end{align}
Taking these relations, we interpolate between them in logarithmic $h/r$ space,  and find that:
\begin{equation}
\delta_{\rm max} = 0.96 \left(\frac{\mplanet/\mj}{(h/r)^{1.38}}\right)^{0.23}.\\
\label{eq:contrast}
\end{equation}
We emphasize that because we only have two points in h/r space, the above relation is only meaningful between $h/r$ = 0.05 and 0.1. It does not directly reflect the physical role $h/r$ plays, and should not be over-interpreted in that sense. As $\delta_{\rm max}-1$ is always a positive number, Equation~\ref{eq:contrast} does not apply to $\mplanet\lesssim(h/r)^{1.38}\mj$ (i.e., very weak waves), although, our data points do show that it can go very close to this limit. The scaling relation \ref{eq:contrast} appears robust, spanning more than two orders of magnitudes in scaled mass in our models. Since the arm constrast in scattered light depends on a variety of factors in a complicated way, this tight correlation is remarkable. We note that the quantity inside the parenthesis in Equation~\ref{eq:contrast} is close to $\sqrt{F}\times\rm constant$, where $F$ is the total torque on one side of the disk exerted by the planet in the linear regime:
\begin{equation}
F \approx \frac{\mplanet^2}{(h/r)^3}\frac{{\rm G}\Sigma_0r}{M_\star}.
\label{eq:f}
\end{equation}
Thus, the scaling in Equation~\ref{eq:contrast} can also be described as $\delta_{\rm max}\propto F^{0.1}$. Surprisingly, the scaling holds very well across $\mplanet\sim\mt$, suggesting linearity may not be an important factor in arm contrast.

Finally, Figure~\ref{fig:contrast_conv} shows $\delta_{\rm max}$ as a function of $\mplanet$ for the convolved images (each radial profile in Figure~\ref{fig:radial_max} contributes a data point). $\delta_{\rm max}$ falls from its value in the full resolution images (Figure~\ref{fig:contrast_raw}), as PSF convolution smears out spiral arms. It is possible to fit a correlation between $\delta_{\rm max}$ and $\mplanet$ for the convolved images, although the scaling is affected by the angular resolution, location of the planet, and distance to the source, so it generally requires a case by case treatment in real observations. 


\section{Discussions}\label{sec:discussions}

\subsection{Inclined Disks}\label{sec:inclined}

In this paper we focus on near-face-on systems. Inclined disks are more complicated for two reasons. First, the shape and strength of an arm can be dramatically affected by the disk inclination and the arm's position angle \citep{dong16armviewing}. Second, a global ``azimuthally averaged'' surface brightness does not exist, as the surface brightness at a given deprojected radius varies with azimuthal angle even without any planet, mainly due to the variation in scattering angle. In general this effect is difficult to correct, due to unknown dust properties. Therefore, Equation~\ref{eq:contrast_r} becomes illy defined, and one can only define local feature contrast by comparing features with their immediate surrounding.

The azimuthal variation of surface brightness caused by inclination in disks with no planets provides an estimate of the minimum mass of planets that can drive distinguishable arms in inclined systems. If the inclination-induced variation matches or exceeds the brightness of the spirals, they will be difficult to detect. As an example, for our assumed dust properties (ISM dust), planet location, and angular resolution, the two horizontal dotted lines in Figure~\ref{fig:contrast_conv} mark the azimuthal variation caused by an inclination $i=40^\circ$ at a projected distance $r=0\arcsec.15$. In the $h/r=0.05$ series, this value is lower than the azimuthal variation caused by a $1\mj$ planet at face-on angle, while in the $h/r=0.1$ series it is lower than the azimuthal variation caused by a $3\mj$ planet. Thus, arms driven by planets with $\mplanet\lesssim1-3\mj$ may be difficult to detect in our disks at 40$^\circ$ inclination.

\subsection{The Effects of Viscosity}\label{sec:viscosity}

In disks, viscous diffusion tends to smear sharp features such as density waves, therefore we expect arm contrast to drop as viscosity increases. In this section we quantify this effect using simulations with finite viscosity. Figure~\ref{fig:viscosity} compares the surface density and scattered light images of three models with $\mplanet=1\mj$, $h/r=0.1$, and different viscosities ($\alpha=0$, $\alpha=10^{-2}$, and $\alpha=10^{-3}$). In full resolution images, the peaks of the primary and secondary arms drop by less than 10\% as $\alpha$ increases from 0 to $10^{-2}$, while the additional arms become significantly weaker in both surface density and in scattered light. The location of the arms are not affected by viscosity. This can be more clearly seen in the azimuthal profiles of the arms (Figure~\ref{fig:viscosity_azimuthal}). Since the global arm contrast is mainly set by the primary/secondary arm, $\delta_{\rm max}$ drops only weakly with increasing $\alpha$ for $\alpha\leq10^{-2}$: in full resolution images $\delta_{\rm max}=2.12$, 2.05, and 1.84 for $\alpha=0$, $10^{-3}$, and $10^{-2}$, while in convolved images $\delta_{\rm max}=1.28$, 1.26, and 1.26 for $\alpha=0$, $10^{-3}$, and $10^{-2}$, respectively.

Although viscosity in protoplanetary disks has been difficult to constrain, recent advances in both theory and observation have hinted that many disks may have low viscosity with $\alpha\lesssim10^{-3}$:
\begin{enumerate}
\item ALMA observations have found that the mm continuum emission in a few transitional disks is lopsided, with most of the emission coming from an azimuthally confined region \citep[e.g.,][]{vandermarel13, casassus13, perez14}. At the moment the most promising explanation for these disks is that the emission is from a vortex formed at the edge of a gap opened by a massive planet, generated by the Rossby wave instability \citep[e.g.,][but also see \citealt{hammer16}]{lin12, lyra13, zhu14votices, zhu14stone}. To realize this scenario, a key ingredient is a low viscosity. \citet{zhu14stone} have shown that $\alpha\lesssim10^{-3}$ is required to form vortices at the gap edge.
\item Non-detections of non-thermal motions induced by disk turbulence in ALMA gas kinematics observations of a few systems suggest low viscosity. For example, based on ALMA CO observations, \citet{flaherty15} determined the level of turbulence in HD~163296 is less than $3\%$ of the local sound speed, implying $\alpha<10^{-3}$.
\item The prime candidate for providing viscosity in disks is the Magnetorotational instability (MRI). The operation of MRI requires the disk to be sufficiently ionized and well coupled to the magnetic field, otherwise MRI cannot be triggered and low viscosity is expected \citep{gammie96}. Recently magnetohydrodynamics (MHD) simulations have shown than non-ideal MHD effects, in particular ambipolar diffusion, can significantly suppress MRI in disks at tens of AU, resulting in very low viscosity equivalent to $\alpha<10^{-3}$ at the bulk of the disk \citep{bai11, bai11stone, bai11grains,perezbecker11-td, perezbecker11, simon13,  turner14, bai15}.
\end{enumerate}

Therefore, in real disks the likely low level of viscosity ($\alpha\lesssim10^{-3}$) will not affect the morphology and contrast of planet-induced arms.

\subsection{How Low Can a Planet's Mass Be to Still Drive Detectable Spiral Arms?}\label{sec:minimum}

In observations, whether a feature is significantly detected or not is determined by its signal-to-noise ratio (S/N). In practice, the noise level in direct imaging of disks at low to mid inclinations can be estimated as the standard deviation of the polar coordinate stokes parameter $U_r=-Q\sin{2\phi}+U\cos{2\phi}$ \citep{schmid06, canovas15qrur}, where $\phi$ is the position angle (see \citealt{thalmann15} for an example). For a given observation, the noise level in physical unit (e.g., mJy arcsec$^{-2}$) in a specific region depends on a variety of factors, including the instrument, the data reduction procedure, the total integration time, the brightness of the star, radial distance, and observing conditions. With today's NIR capability the noise level can often get down to $\sim1-{\rm a\ few} \times0.1$~mJy arcsec$^{-2}$ \citep[e.g.,][]{hashimoto12, stolker16, ohta16}. 

Assessing the detectable planet mass limit generally requires a case by case effort. Here we provide an example to show how this can be done. If we specify a $3\sigma$ detection threshold as 1 mJy arcsec$^{-2}$ at the location of our inner arms, Figure~\ref{fig:azimuthal_conv} shows that in convolved images, spiral arms driven by Saturn mass planets in our models are marginally detectable (convolved images of the 1MS-H5 model can be found in Figure~\ref{fig:image_tmt}). These detection thresholds are affected by several factors. A finer angular resolution (or smaller source distance) improves the detection threshold by bringing the arm contrast in convolved images closer to in full resolution images. A brighter star on the other hand raises both the disk surface brightness and the noise level; whether that improves or deteriorates S/N depends on the scaling. For planets at larger radii, a smaller physical resolution under the same angular resolution makes the arms sharper (higher contrast), but their intrinsic surface brightness drops due to larger distance from the star. Case-by-case modeling is needed to derive the minimum detectable planet mass with different distance, angular resolution, disk $h/r$, planet location, and stellar luminosity.

\subsection{Comments on the Currently Observed List of Spiral Arms}\label{sec:minimum}

To date, a number of systems have been identified to have spiral arms in scattered light\footnote{Not all spiral arms are density waves excited by planets. For example, mirror-symmetric pairs of arms in modest-to-high inclination systems like AK Sco are likely to be part of a disk ring, \citet{janson16}, \citet{dong16armviewing}.}. Two distinct classes exist:
\begin{enumerate}[(A)]
\item MWC~758, SAO~206462, HD~100453, and LkH$\alpha$~330, which show large scale arms in pairs with $\delta_{\rm max}\gtrsim2$, azimuthal extension $\gtrsim\pi$, and a near $m=2$ rotational symmetry.
\item HD 142527 and AB Aur, which show half a dozen or more small scale arms with $\delta_{\rm max}\lesssim1.5$, azimuthal extension $\lesssim\pi/2$, and no obvious rotational symmetry. 
\end{enumerate}
The exceptions are HD 100546, which shows very complicated structures \citep[e.g.,][]{currie15, garufi16}, with the two main arms possibly being part of a disk ring viewed at modest inclination as in AK Sco, and V1247, which shows only one but high contrast arm \citep{ohta16}.

For both classes, it is very unlikely that each arm originates from a different planet; instead, the number of arms is likely larger than the number of arm-driving planets. Our simulations show that even a 3 Neptune mass planet in a hot disk ($h/r=0.1$; left panel in Figure~\ref{fig:3mn1mj3mj}) and a 1 Neptune mass planet in a cold disk ($h/r=0.05$; top row in Figure~\ref{fig:image_tmt}) drive multiple arms. Moreover, these arms are weak, with $\delta_{\rm max}\sim1.2$ in full resolution images (Figure~\ref{fig:contrast_raw}). When convolved by PSFs and with observational noises added, these arms will likely be either undetectable, or comparable to the weakest detected arms so far (e.g., AB Aur). For a single planet to drive only one arm, the planet mass has to be lower than 1-3 Neptune masses, and such arms are very unlikely to be detectable with current NIR capability. Based the \citet{fung15} study of the symmetry of the arms driven by one companion, the pairs of arms in class (A) are likely produced by a single massive perturber on the order of 10 $\mj$ or more outside the arms, while the arm-exciting planets in class (B) are likely to be on the order of Jupiter mass. Note that in the former cases, the predicted external perturbers would carve out detectable gaps around their orbits (and if given a sufficiently long time the planets in class (B) systems may do so as well).

Furthermore, class (A) arms cannot be fit with the linear density wave theory, as has often been assumed in the literature. Figure~\ref{fig:contrast_raw} shows that even in the limit of full resolution images (i.e. maximum contrast), to reach $\delta_{\rm max}=2$, the planet mass still has to be $\mplanet\sim\mt$ in a hot disk ($h/r=0.1$), and several $\mt$ in a cold disk ($h/r=0.05$). \citet{zhu15densitywaves} showed that with $\mplanet\gtrsim\mt$, planet-induced spiral arms are in the fully nonlinear regime and have larger pitch angle than predicted by the linear theory.

Lastly, we note that the arm directly imaged in the V1247 Orionis system \citep{ohta16} has an astonishingly high contrast; $\delta_{\rm max}$ at the project distance of the arm is at least a factor of $\sim3$. Although this disk has a non-negligible inclination $\sim30^\circ$, our results still suggest that the perturber in this case is likely to be massive (several Jupiter mass or above). At least two arms are expected, while the second arm may be under the HiCIAO mask, because in inclined systems the arms excited by a single companion can be at quite different radial distances due to viewing angle effects \citep{dong16armviewing}.

\subsection{Application to SAO 206462}\label{sec:sao}

As an application of our modeling results, we examine the arms in the SAO 206462 disk (central star mass 1.7 $\msun$; distance 140 pc; \citealt{muller11, vanboekel05}). The inclination of the system, $i\sim10-20^\circ$ constrained by mm observations \citep[e.g.,][]{lyo11, perez14, vandermarel15}, is among the lowest in all spiral arm disks. The system is close enough to being face-on so that approximately our model results at $i=0$ apply. By modeling the SED and the mm dust continuum emission of the system, \citet{andrews11} determined $h/r=0.096$ at 100 AU in the disk with a flaring index of 0.15. \citet{dong15spiralarms} predicted that a planet, SAO 206462 b, outside the two arms at $\sim0\arcsec.7\approx100$ AU from the center, can explain the general morphology of the pair of arms. Thus, by rescaling our $h/r=0.1$ models to put the planet at 100 AU, we can constrain the mass of SAO 206462 b based on its arm contrast. Note that our models are more flared (flaring index 0.25) than the \citet{andrews11} fitting outcome. Nevertheless, we do not expect the arm contrast to be significantly different.

Figure~\ref{fig:sao} shows the results. We measure $\delta(r)$ in the VLT/NACO $Ks$-band dataset \citep{garufi13} based on Equation~\ref{eq:contrast_r}. For a planet at 100 AU, $\delta(r)$ is expected to peak around 60--70 AU. In reality, $\delta(r)$ in SAO 206462 stays within 1.8--2 in between 53--72 AU, before dropping toward smaller radius (beyond $\sim75$ AU the disk is too faint to perform a robust $\delta(r)$ measurement). We rescale our $h/r=0.1$ models to have $\rp$=100 AU, and convolve the full resolution images to reach an angular resolution of $0\arcsec.09$ to match the observation \citep{garufi13}. The $\delta(r)-\mplanet$ curve is shown on the right panel. Comparing SAO 206462's arm contrast with models suggests $q\sim3-6\times10^{-3}$, or $\mplanet\sim5-10\mj$ given $M_\star=1.7\msun$. This is broadly consistent with the estimate $\mplanet\sim6\mj$ made by \citet{fung15}, based on the relation between the planet mass and the azimuthal separation of the two arms.

\subsection{Imagining Planet-Driven Spiral Arms with Thirty-Meter-Class Telescopes}\label{sec:tmt}

Several thirty meter class telescopes (TMCT for short) are planned, such as the Thirty Meter Telescope, the European Extremely Large Telescope, and the Giant Magellan Telescope. Although it is still too early to know the science definitions of the imaging instruments that will be commissioned on these telescopes, one thing for certain is that they will greatly improve the angular resolution in direct imaging, governed by the $\lambda/D$ law. A factor of 3 to 4 improvement the over current level achieved by GPI and SPHERE will dramatically push down the detectable mass limit of arm-driving planets. Figure~\ref{fig:image_tmt} compares synthetic images of arms driven by a Neptune and a Saturn mass planet at 30 AU convolved by the currently available $0\arcsec.04$ PSF and a $0\arcsec.01$ PSF achievable by TMCT. Clearly, with $0\arcsec.04$ PSF the arms driven by a Saturn mass planet are only marginally recognizable out to a distance of 140 pc, while the arms driven by a Neptune planet are hopelessly too faint, but with the $0\arcsec.01$ PSF, even a Neptune in a solar-nebular ($\rp=30$ AU; $h/r\sim0.05$) drive clearly visible arms.


\section{Summary}\label{sec:summary}

Using 3D hydrodynamics and radiative transfer simulations, we quantitatively examine the morphology and contrasts of planet-induced arms in protoplanetary disks. The range of planet mass in our models spans more than one order of magnitude above and blow the disk thermal mass. Our main conclusions are (the quotes on planet masses below assumes a $1\msun$ central star):
\begin{enumerate}
\item A numerical fitting function is obtained for the maximum arm-to-disk brightness contrast $\delta_{\rm max}$ as functions of $\mplanet$ and $h/r$ (Equation~\ref{eq:contrast}). The empirical relation suggests the arm contrast is a shallow power law of the planetary torque $\mplanet^2/(h/r)^3$.
\item With an angular resolution of $0\arcsec.04$ and at 140 pc, the minimum mass of a planet at 30 AU that drives detectable spiral arms in a disk around a 1 Myr 1 $\msun$ star at face-on view is about Saturn mass (Section~\ref{sec:minimum}). With a factor of 4 improvement in angular resolution enabled by a thirty meter class telescope, the detection threshold may be pushed down to at least Neptune mass (Section~\ref{sec:tmt}).
\item In observed spiral arm systems so far (e.g., MWC~758, SAO~206462, HD~100453, LkH$\alpha$~330, HD 142527, and AB Aur), it is very unlikely that each arm originates from a different planet, because the planets that only drive one arm must be below several Neptune masses, and their spirals arms are most likely not dectectable (Section \ref{sec:minimum}).
\item Similarly, the prominent arms observed in MWC~758, SAO~206462, and LkH$\alpha$~330 cannot be fit with the linear theory \citep[][Equation 44]{rafikov02}, as commonly assumed, because the high contrasts of these arms (as well as their symmetry; \citealt{fung15}) suggest that the perturbers must have $\mplanet>\mt$ (Figure \ref{fig:contrast_raw}), thus the wave excitation and propagation are in the fully nonlinear regime \citep{zhu15densitywaves,lee16}.
\item The contrast of the arms in the SAO 206462 disk suggests that the perturber SAO 206462 b at $\sim100$ AU is about $5-10\mj$ in mass. This is consistent with the previous \citet{fung15} estimate based on arm morphology (Section~\ref{sec:sao}).
\item The contrast of the outer arms is lower than the inner arms except when $\mplanet\gtrsim$a few$\times\mt$ (Figure~\ref{fig:radial_max}), consistent with what was found by \citet{dong15spiralarms}. Combining with the fact that the outer disk is intrinsically fainter, observed arms so far are most likely to be inward of the planet's orbit.
\item We predict that at least one other arm, currently under the HiCIAO mask, should exist in the V1247 Orionis system \citep{ohta16} if the currently detected one is planet-induced, because the strength of the arm suggests a massive perturber. Future imaging observations with smaller inner working angle may be able to detect it.
\item We extend the previous work by \citet{zhu15densitywaves} to examine synthetic NIR images of spiral arms produced by puffing up 2D hydro surface density maps, and compare them with synthetic images directly based on 3D hydro simulations. The inner arms in the formal are significantly weaker than in the latter, due to the missing of vertical kinematic support (Figure~\ref{fig:2d3d}). Thus, 3D hydro simulations are needed in order to study the morphology of planet-induced arms.
\item Finally, the morphology and contrast of the primary and secondary arms are largely unaffected by a modest level of viscosity of $\alpha\lesssim0.01$, while the additional arms are significantly weakened by an $\alpha=0.01$ viscosity (Figures~\ref{fig:viscosity} and \ref{fig:viscosity_azimuthal}).
\end{enumerate}


\section*{Acknowledgments}

We thank Eugene Chiang, Misato Fukagawa, Sascha Quanz, Tomas Stolker, and Gaspard Duchene for helpful discussions, and Antonio Garufi and Sascha Quanz for kindly sharing with us the VLT/NACO dataset of SAO 206462. We also thank the anonymous referee for constructive suggestions that improved the quality of the paper. This project is partially supported by NASA through Hubble Fellowship grant HST-HF-51320.01-A (R.D.) awarded by the Space Telescope Science Institute, which is operated by the Association of Universities for Research in Astronomy, Inc., for NASA, under contract NAS 5-26555. J.F. gratefully acknowledges support from the Natural Sciences and Engineering Research Council of Canada, the Center for Integrative Planetary Science at the University of California, Berkeley, and the Sagan Fellowship Program under contract with the Jet Propulsion Laboratory (JPL) funded by NASA and executed by the NASA Exoplanet Science Institution. This research used the SAVIO computational cluster at UC Berkeley, managed by the IT Division at the Lawrence Berkeley National Laboratory (Supported by the Director, Office of Science, Office of Basic Energy Sciences, of the U.S. Department of Energy under Contract No. DE-AC02-05CH11231); we particularly acknowledge the help from Yong Qin and Krishna Muriki.


\clearpage
\begin{table}[]
\centering
\label{tab:models}
\begin{tabular}{c|cccc}
\hline
Model Name & $h/r$ & $\mplanet$ & $\alpha$ \\ \hline
1MN-H5  & 0.05 & $1\mn$ & 0    \\
3MN-H5  & 0.05 & $3\mn$ & 0    \\
1MS-H5  & 0.05 & $1\ms$ & 0    \\
1MJ-H5  & 0.05 & $1\mj$ & 0    \\
3MJ-H5  & 0.05 & $3\mj$ & 0    \\
1MN-H10 & 0.1 & $1\mn$ & 0    \\
3MN-H10 & 0.1 & $3\mn$ & 0    \\
1MS-H10 & 0.1 & $1\ms$ & 0    \\
1MJ-H10 & 0.1 & $1\mj$ & 0    \\
3MJ-H10 & 0.1 & $3\mj$ & 0    \\
\hline
1MJ-H10-A3 & 0.1 & $1\mj$ & $10^{-3}$    \\
1MJ-H10-A2 & 0.1 & $1\mj$ & $10^{-2}$    \\
\hline
\end{tabular}
\caption{Model List. The first 10 simulations are inviscid, and the last two are viscous. $\rp=30$~AU in all cases. For the $h/r=0.05$ series, $\mt=0.125\mj$; for the $h/r=0.1$ series, $\mt=1\mj$.}
\end{table}

\begin{figure}
\begin{center}
\includegraphics[trim=0 0 0 0, clip,width=0.5\textwidth,angle=0]{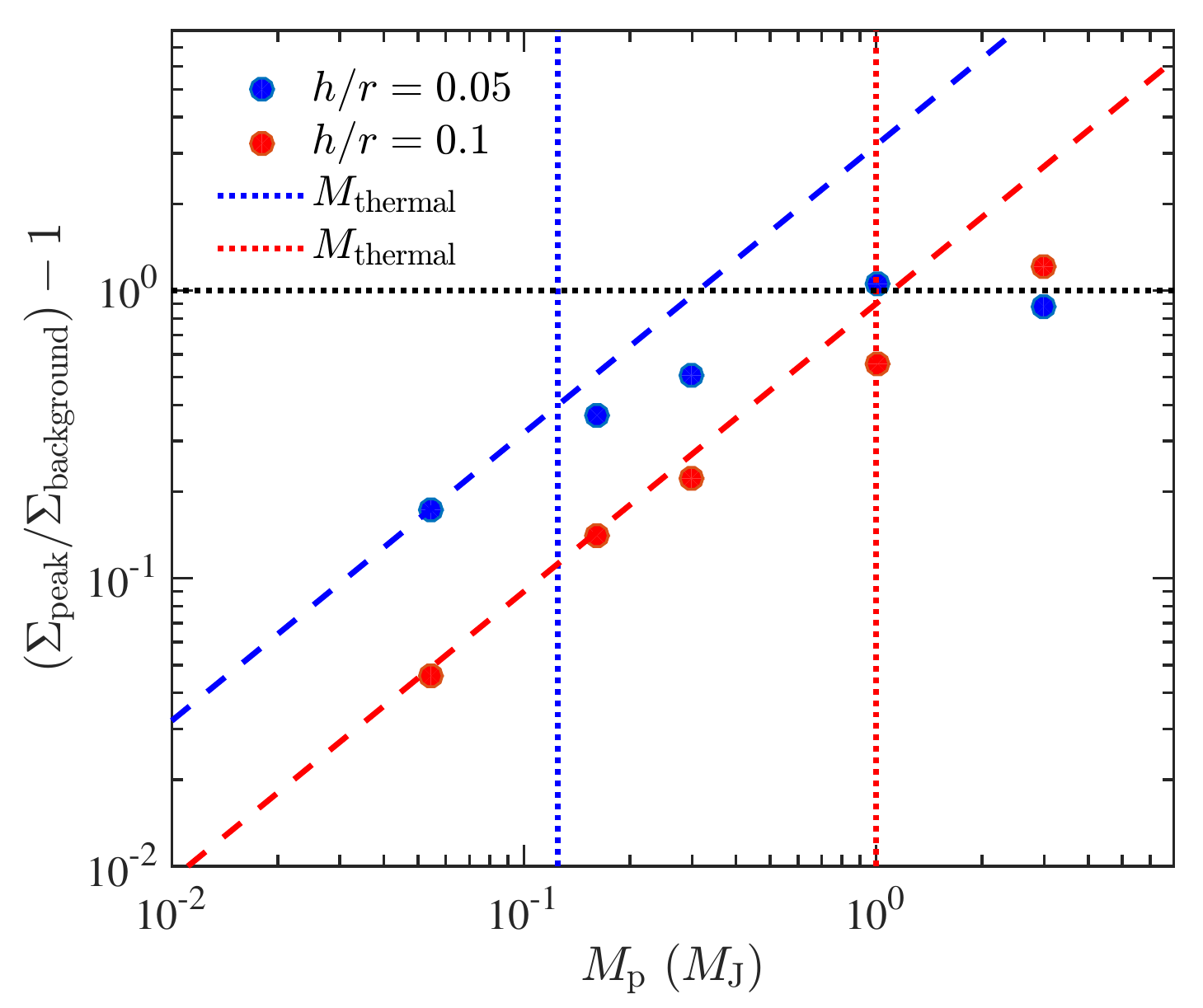}
\end{center}
\figcaption{Surface density perturbation in our models at $r=0.7\rp=21$ AU. Blue symbols and lines are for the $h/r=0.05$ series and red symbols and lines are for the $h/r=0.1$ series. $\Sigma_{\rm peak}$ is the peak surface density at the location of the arms and $\Sigma_{\rm background}$ is the azimuthally averaged surface density. The two vertical dotted lines indicate the thermal mass for each $h/r$ series ($\mt=1\mj$ for $h/r=0.1$ and $\mt=0.125\mj$ for $h/r=0.05$), and the two dashed lines indicate the analytical trend $(\Sigma_{\rm peak}/\Sigma_{\rm background})-1\propto\mplanet$ for low mass planets with $\mplanet\lesssim\mt$ \citep{goodman01}. For massive planets, density perturbation saturates at order unity (indicated by the horizontal dotted line).
\label{fig:sigma_contrast}}
\end{figure}

\begin{figure}
\begin{center}
\includegraphics[trim=0 0 0 0, clip,width=\textwidth,angle=0]{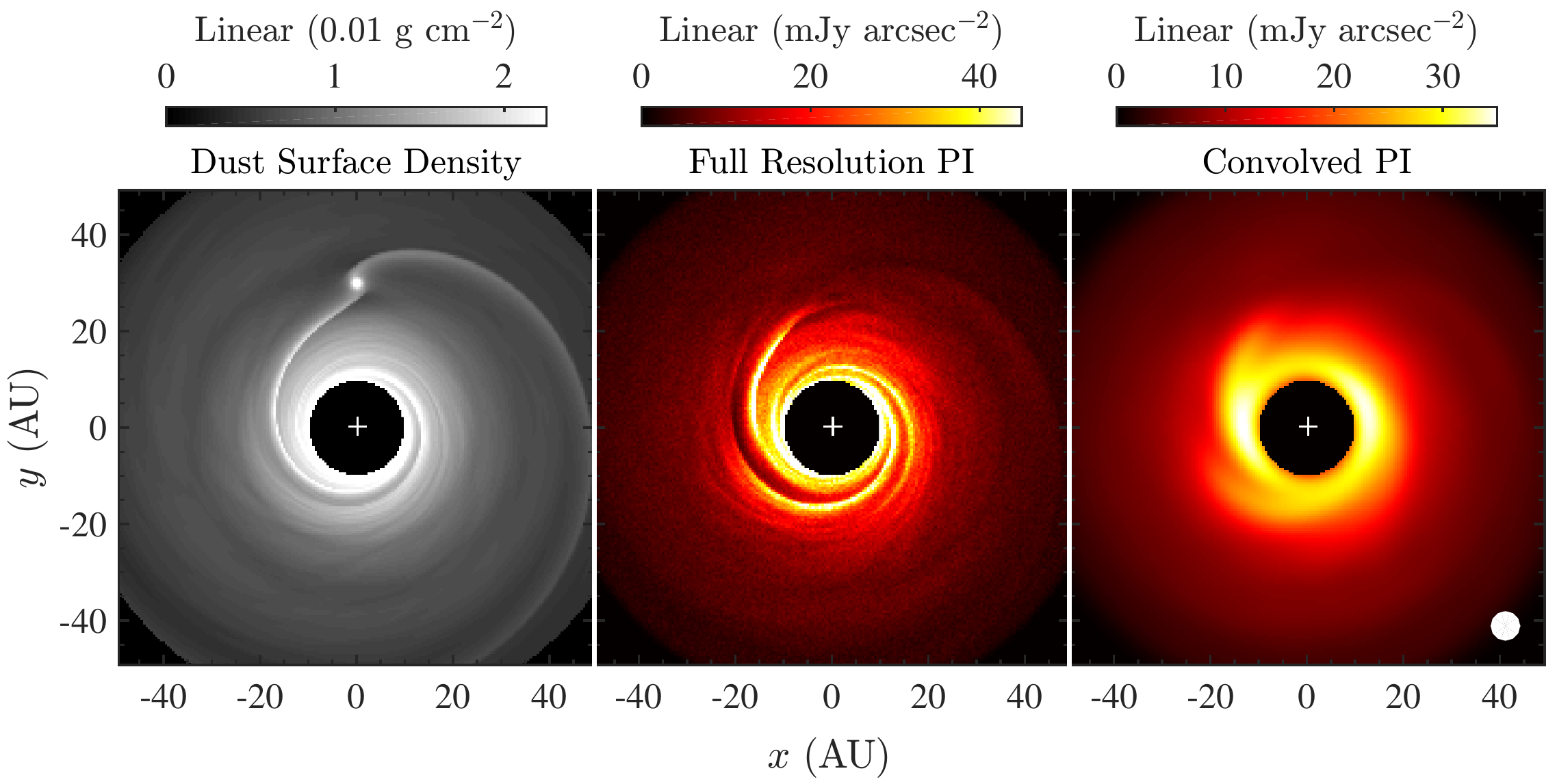}
\end{center}
\figcaption{A basic set of simulation results for the 1MJ-H10 model. {\bf From left to right}: surface density of the dust, full resolution polarized intensity image at $H$-band, and convolved image (the $0\arcsec.04$ Gaussian PSF size is marked at the lower right corner). The location of the planet is at [0,30]~AU, evident in the surface density map. The ``+'' sign at the center marks the location of the star. The central $0\arcsec.07$ is blocked to mimic the effect of an inner working angle.
\label{fig:basic3}}
\end{figure}

\begin{figure}
\begin{center}
\includegraphics[trim=0 0 0 0, clip,width=0.8\textwidth,angle=0]{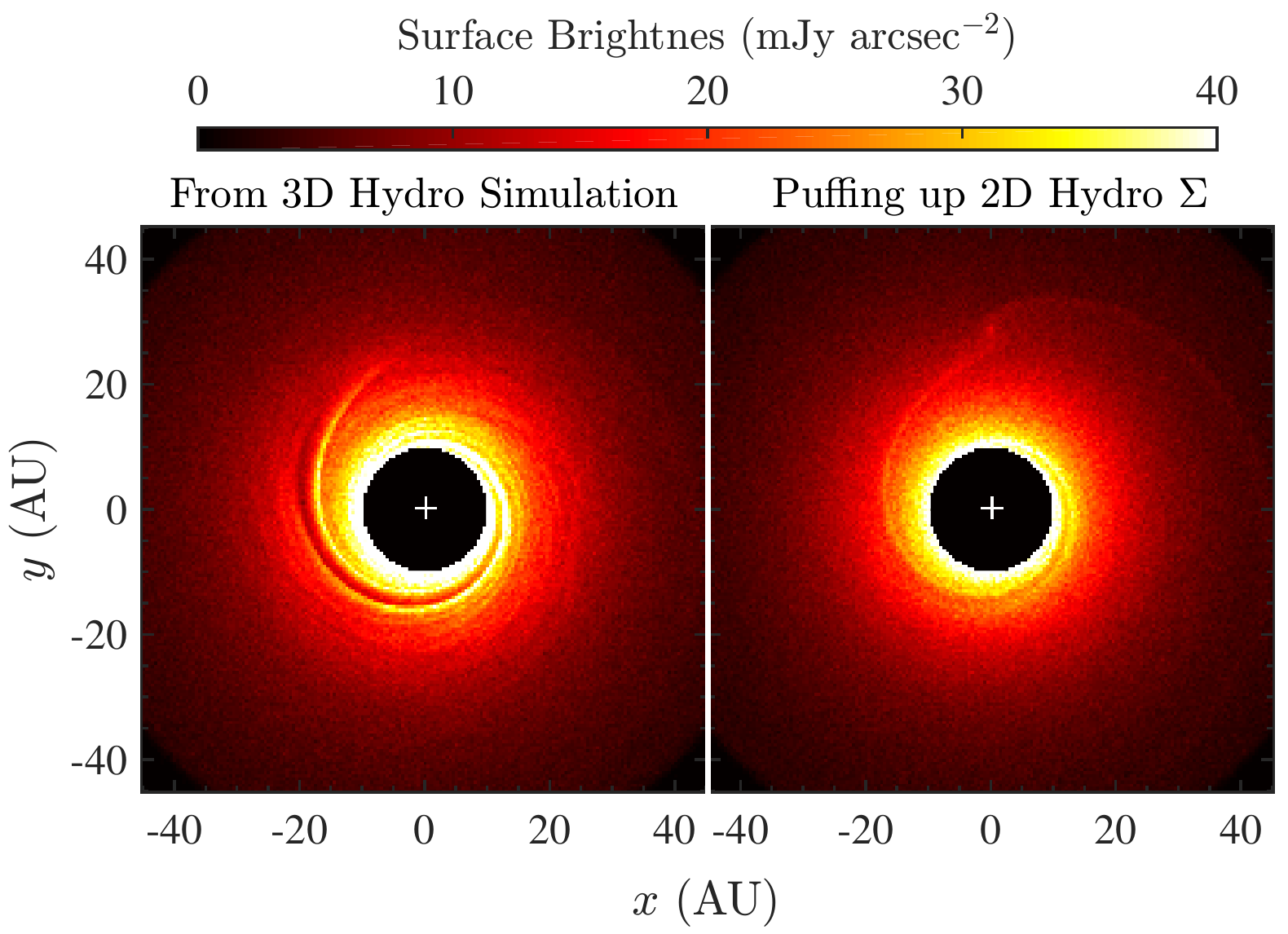}
\includegraphics[trim=0 0 0 0, clip,width=0.5\textwidth,angle=0]{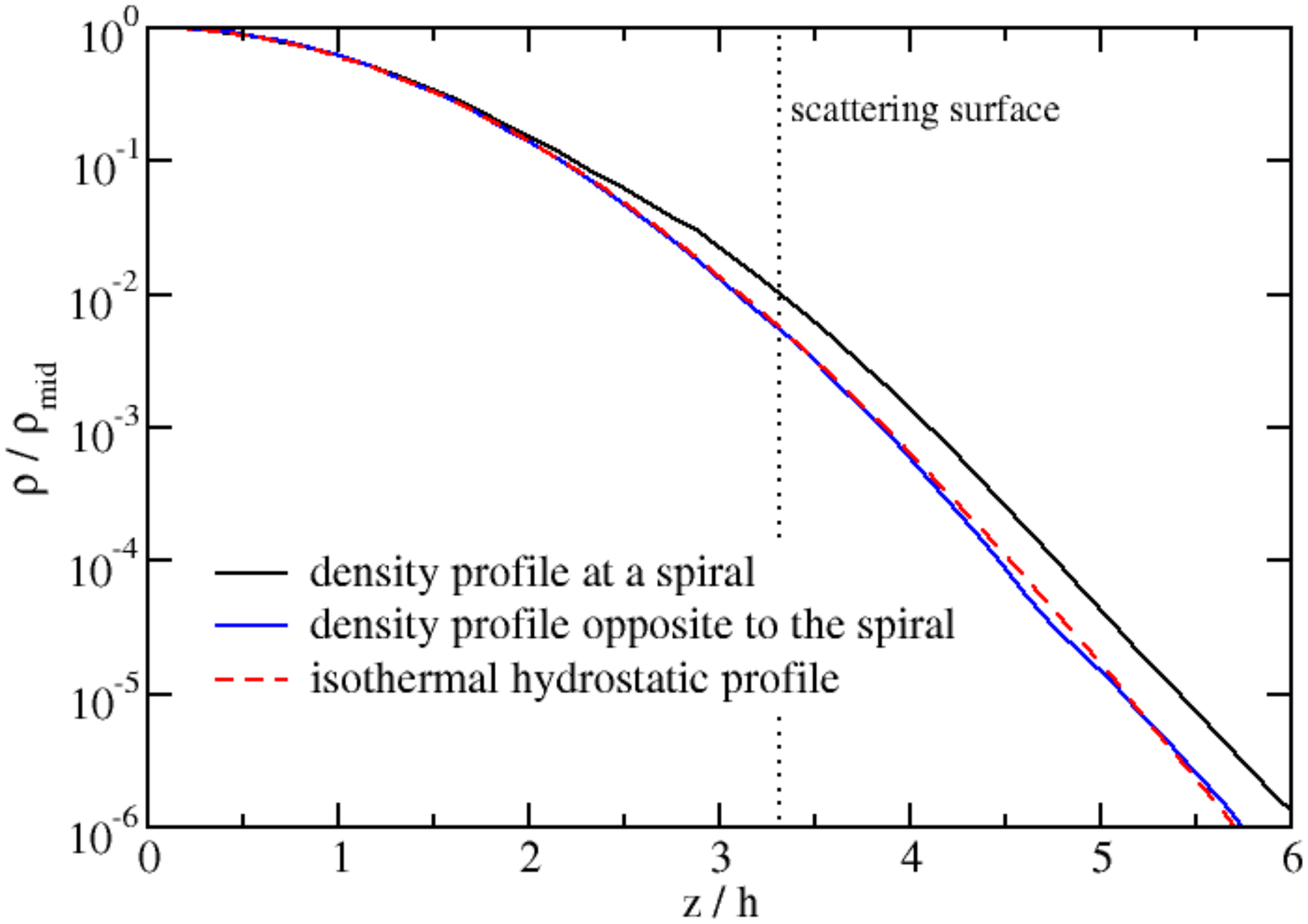}
\end{center}
\figcaption{{\bf Top:} Synthetic images based on full 3D hydro simulation ({\bf left}) and hydro surface density puffed up assuming hydrostatic equilibrium in the vertical direction ({\bf right}; with the input hydro $h/r$) for the 1MS-H10 model. The waves on the right panel is significantly less visible than on the left panel. {\bf Bottom:} Comparison between the vertical density profiles in the 3D hydro simulation and the hydrostatic solution (Equation \ref{eqn:rho_vertical}). The black curve shows the vertical density profile at a spiral; the blue curve shows the profile on the opposite side in the disk at the same radial distance; and the red dashed curve is the isothermal hydrostatic profile. The vertical dotted line indicates the location of the scattering surface at this location (the radially integrated optical depth at $H$-band between the star and this surface is unity). When moving away from the disk midplane, density decreases slower in the arm in 3D simulations, due to vertical kinematic support. See Section~\ref{sec:2d3d} for details.
\label{fig:2d3d}}
\end{figure}

\begin{figure}
\begin{center}
\includegraphics[trim=0 0 0 0, clip,width=\textwidth,angle=0]{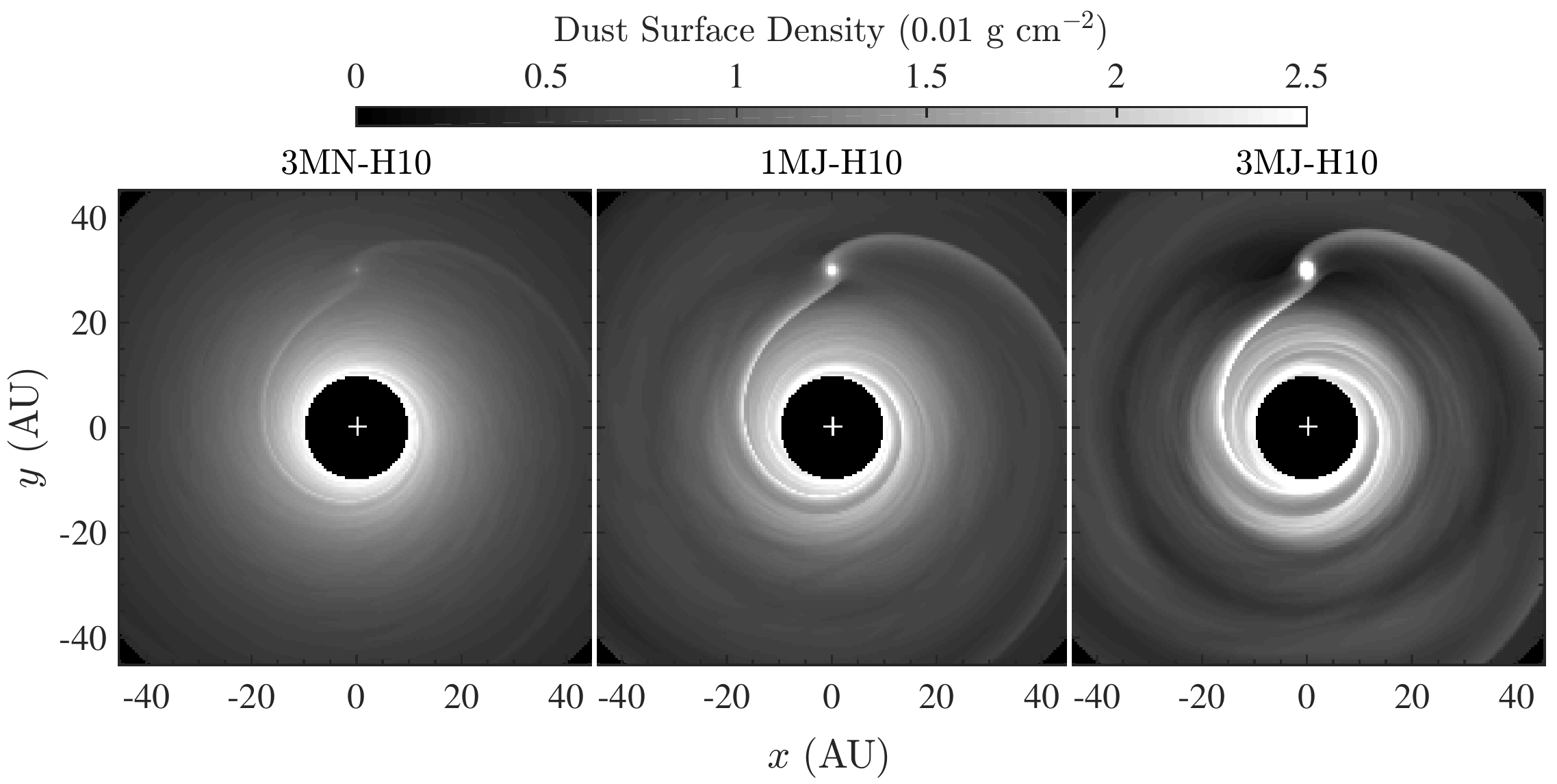}
\includegraphics[trim=0 0 0 0, clip,width=\textwidth,angle=0]{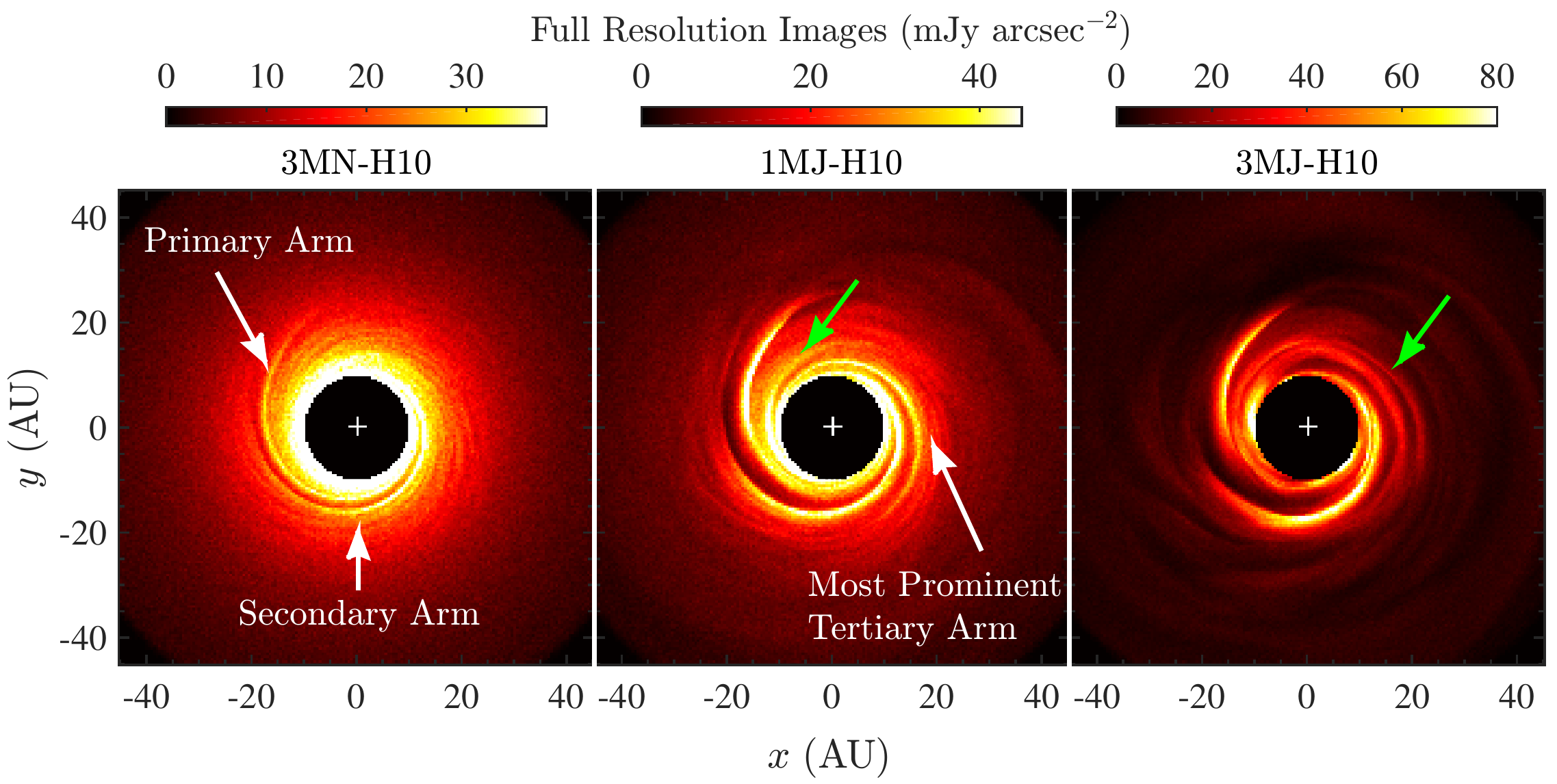}
\end{center}
\figcaption{Surface density ({\bf top row}) and full resolution $H$-band PI images ({\bf bottom row}) of 3 models with $h/r=0.1$. {\bf From left to right}: $3\mn=0.16\mt$, $1\mj=1\mt$, and $3\mj=3\mt$. The primary and the secondary arms are visible in all scattered light images and have roughly equal strengths, while the tertiary is only visible in the latter two cases with $\mplanet\gtrsim\mt$, while achieving a similar strengths with the primary/secondary only in the 1MJ-H10 model. The secondary in 3MN-H10 and the tertiary in 1MJ-H10 are very weak in the surface density, thus their prominence in images must originate from vertical kinematic support. Additional features that may be considered as a $4^{\rm th}$ (and more) arm can also be found in the $1\mt$ and $3\mt$ cases, indicated by green arrows. See Section~\ref{sec:arms} for details.
\label{fig:3mn1mj3mj}}
\end{figure}

\begin{figure}
\begin{center}
\includegraphics[trim=0 0 0 0, clip,width=0.49\textwidth,angle=0]{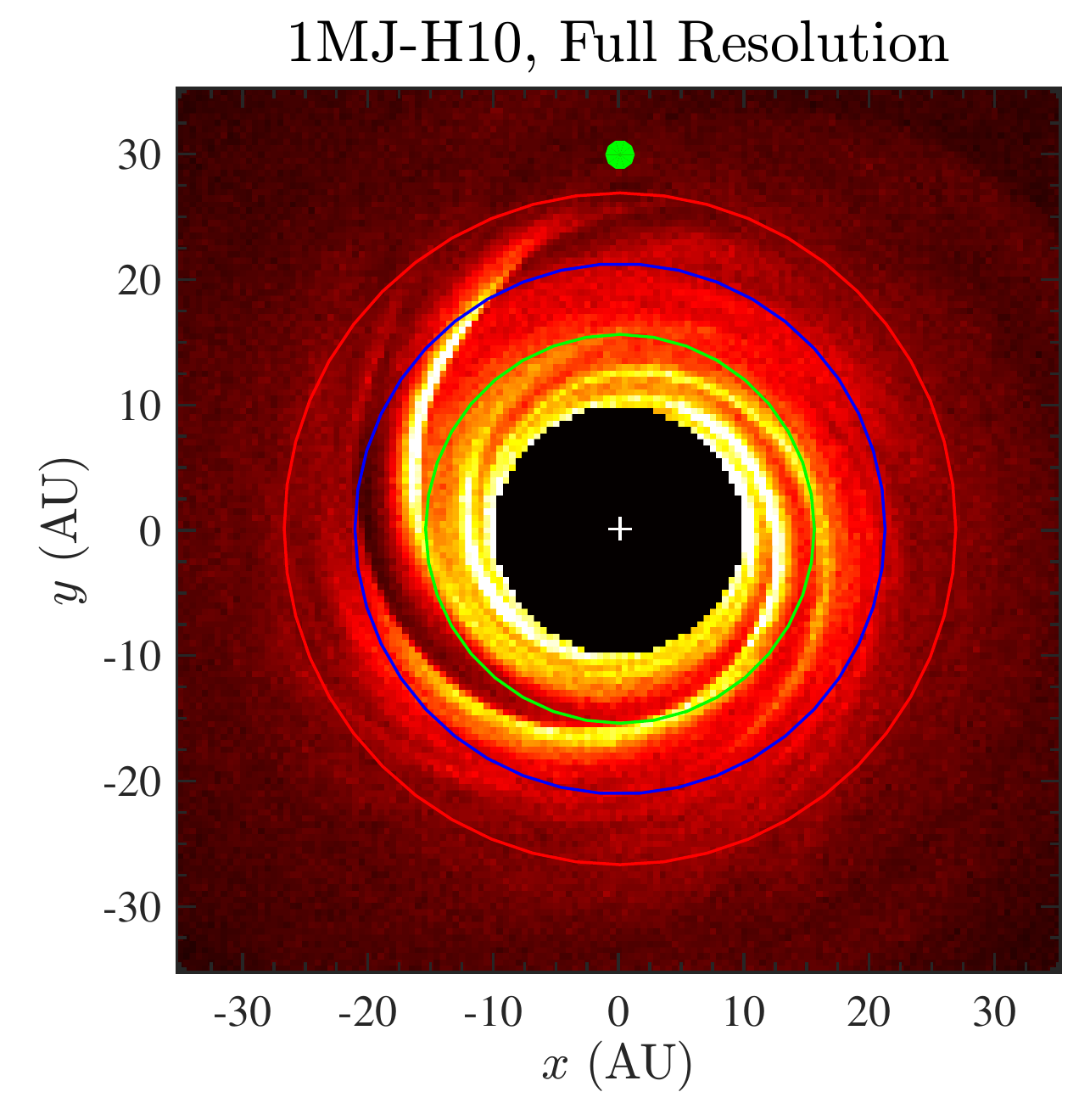}
\includegraphics[trim=0 0 0 0, clip,width=0.49\textwidth,angle=0]{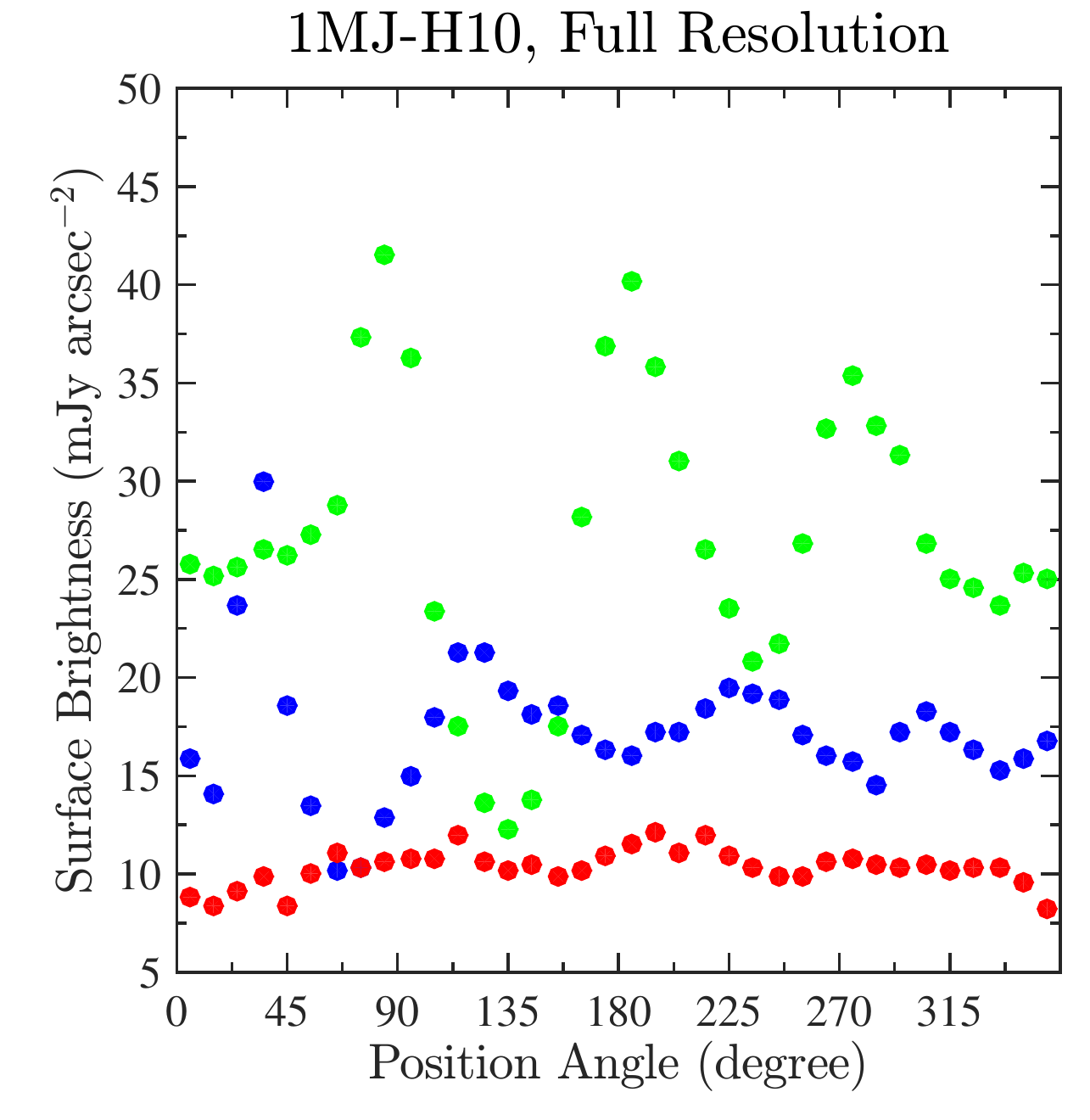}
\includegraphics[trim=0 0 0 0, clip,width=0.49\textwidth,angle=0]{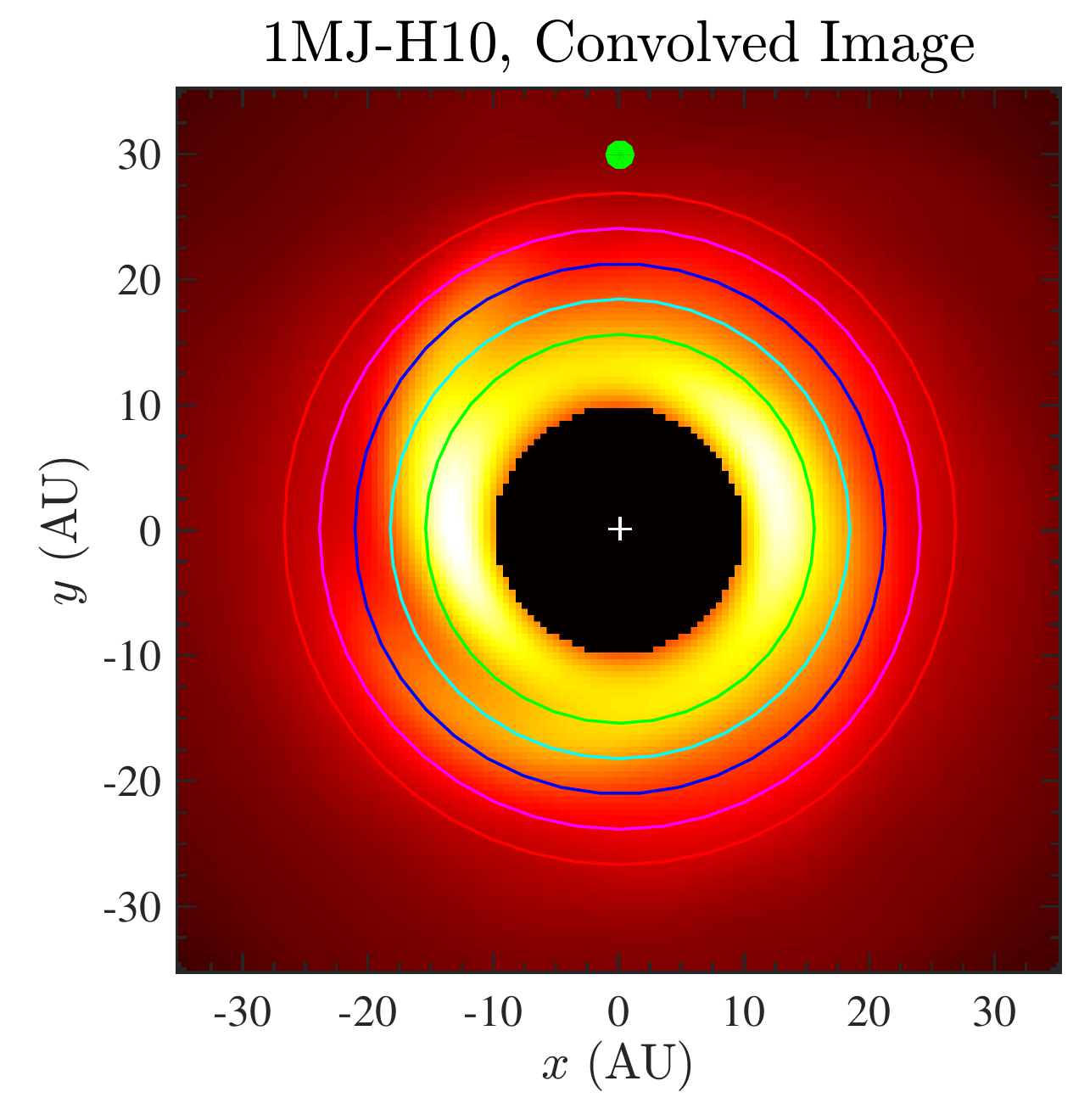}
\includegraphics[trim=0 0 0 0, clip,width=0.49\textwidth,angle=0]{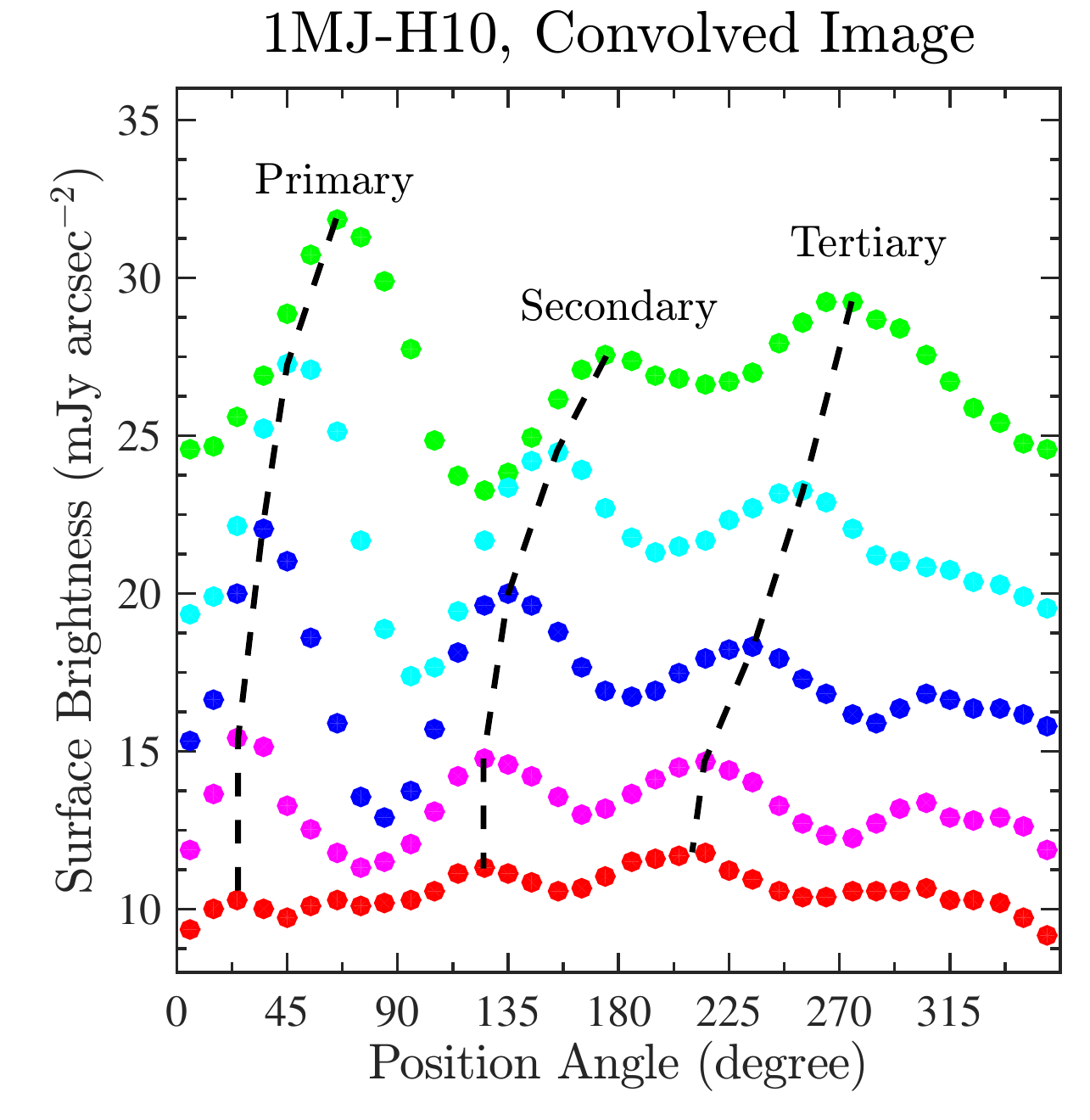}
\end{center}
\figcaption{{\bf Top row:} full resolution image of the 1MJ-H10 model ({\bf left}) and its azimuthal profile ({\bf right}) at $r=0\arcsec.11,0\arcsec.15,0\arcsec.19$ (15.4, 21, 26.6 AU, indicated by the circles with the same color on the left panel), from top to bottom. {\bf Bottom row:} the same as the top row, but for the convolved image, and the profiles are shown for $r=0\arcsec.11,0\arcsec.13,0\arcsec.15,0\arcsec.17,0\arcsec.19$ (color scheme remains the same). The position angle (PA) is measured counterclockwisely from North (thus the planet, marked by the green dot, is at PA=$0^\circ$). The 3 dashed lines in the bottom right trace out the three major arms. The azimuthal separations between the arms stay roughly constant. See Section~\ref{sec:contrast} for details.
\label{fig:azimuthal_1mj_h10}}
\end{figure}

\begin{figure}
\begin{center}
\includegraphics[trim=0 0 0 0, clip,width=\textwidth,angle=0]{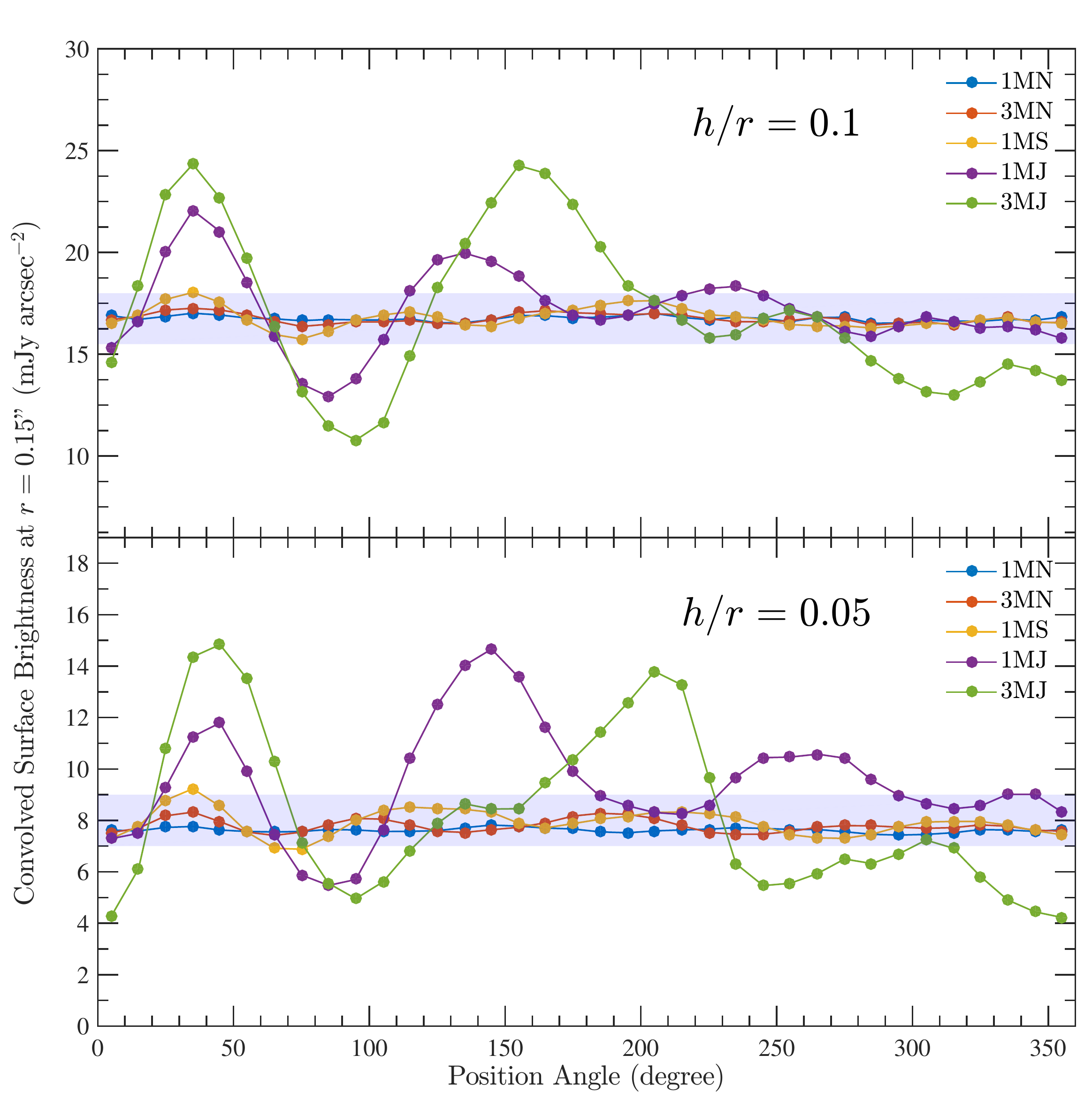}
\end{center}
\figcaption{Azimuthal profiles of convolved images at $r=0\arcsec.15$ (21 AU) for the $h/r=0.1$ (top) and $h/r=0.05$ (bottom) series. The shaded horizontal band in each panel marks $\pm1$ mJy arcsec$^{-2}$ around the average, which roughly indicates the detection threshold (3$\sigma$ noise level) of features in current NIR polarized scattered light imaging observations. See Section~\ref{sec:contrast} for details.
\label{fig:azimuthal_conv}}
\end{figure}

\begin{figure}
\begin{center}
\includegraphics[trim=0 0 0 0, clip,width=0.49\textwidth,angle=0]{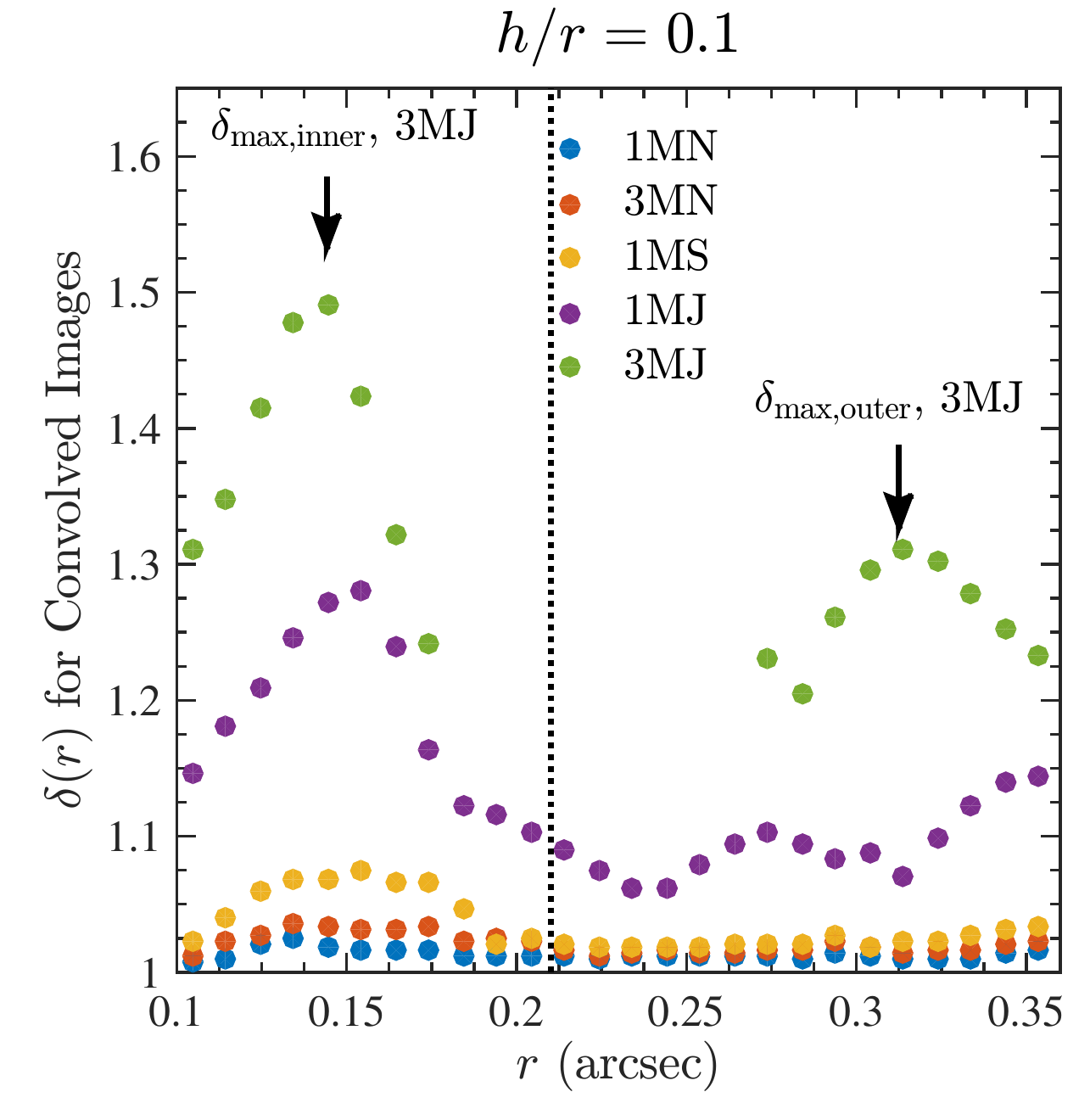}
\includegraphics[trim=0 0 0 0, clip,width=0.49\textwidth,angle=0]{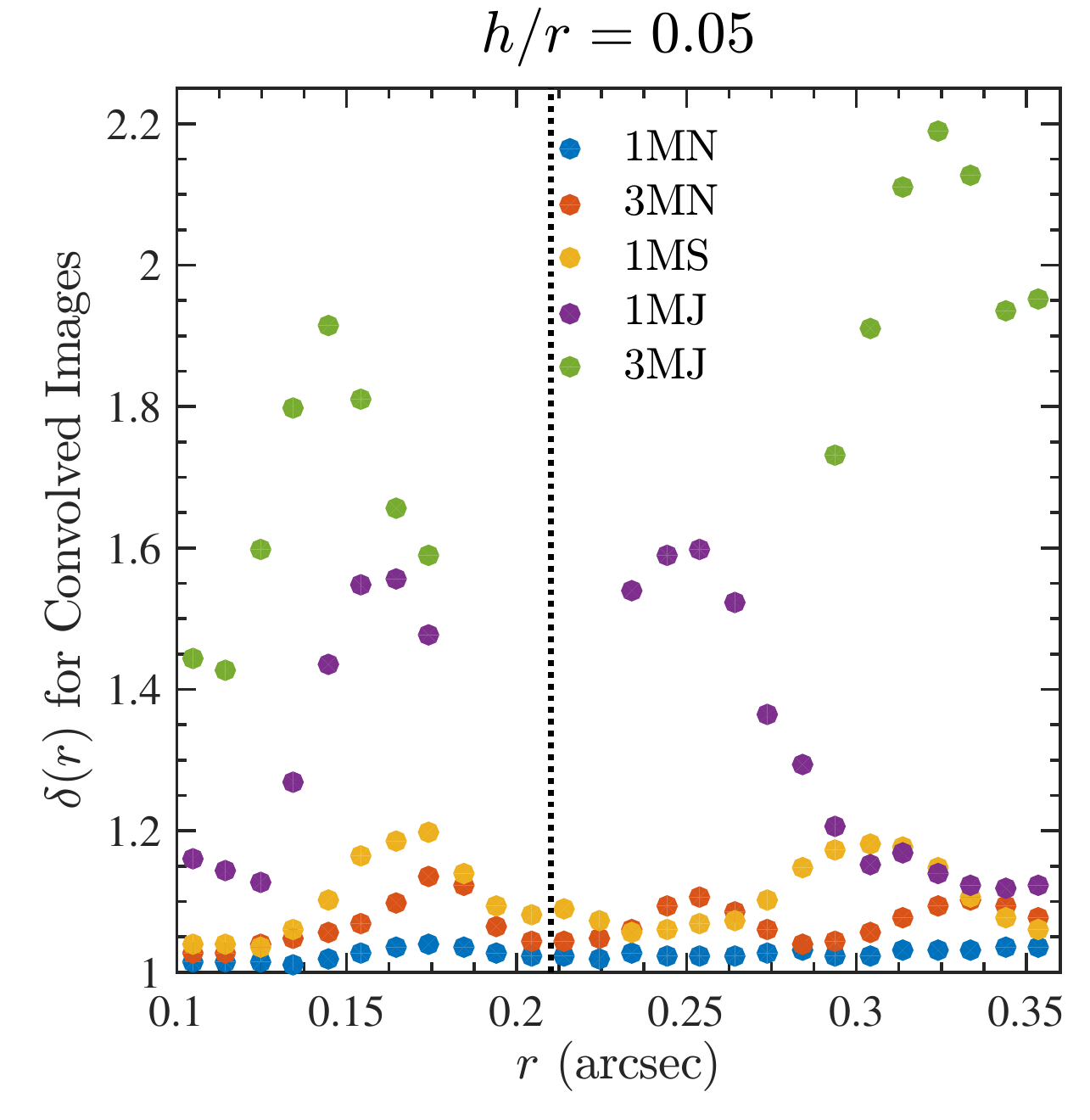}
\end{center}
\figcaption{The contrast of the arms at each radius, defined in Equation~\ref{eq:contrast_r}, for convolved images (each azimuthal profile in Figures~\ref{fig:azimuthal_conv} and \ref{fig:azimuthal_1mj_h10} contributes one data point). The {\bf left} and {\bf right} panels are for the $h/r=0.1$ and $h/r=0.05$ series, respectively. The 1MJ-H5, 3MJ-H5, and 3MJ-H10 models start to carve out a noticeable gap even at 10 orbits; the regions affected by the gaps are not shown to avoid confusion (arms in gaps are not well defined). The vertical dotted line indicates the orbital radius of the planet. See Section~\ref{sec:contrast} for details.
\label{fig:radial_max}}
\end{figure}

\begin{figure}
\begin{center}
\includegraphics[trim=0 0 0 0, clip,width=\textwidth,angle=0]{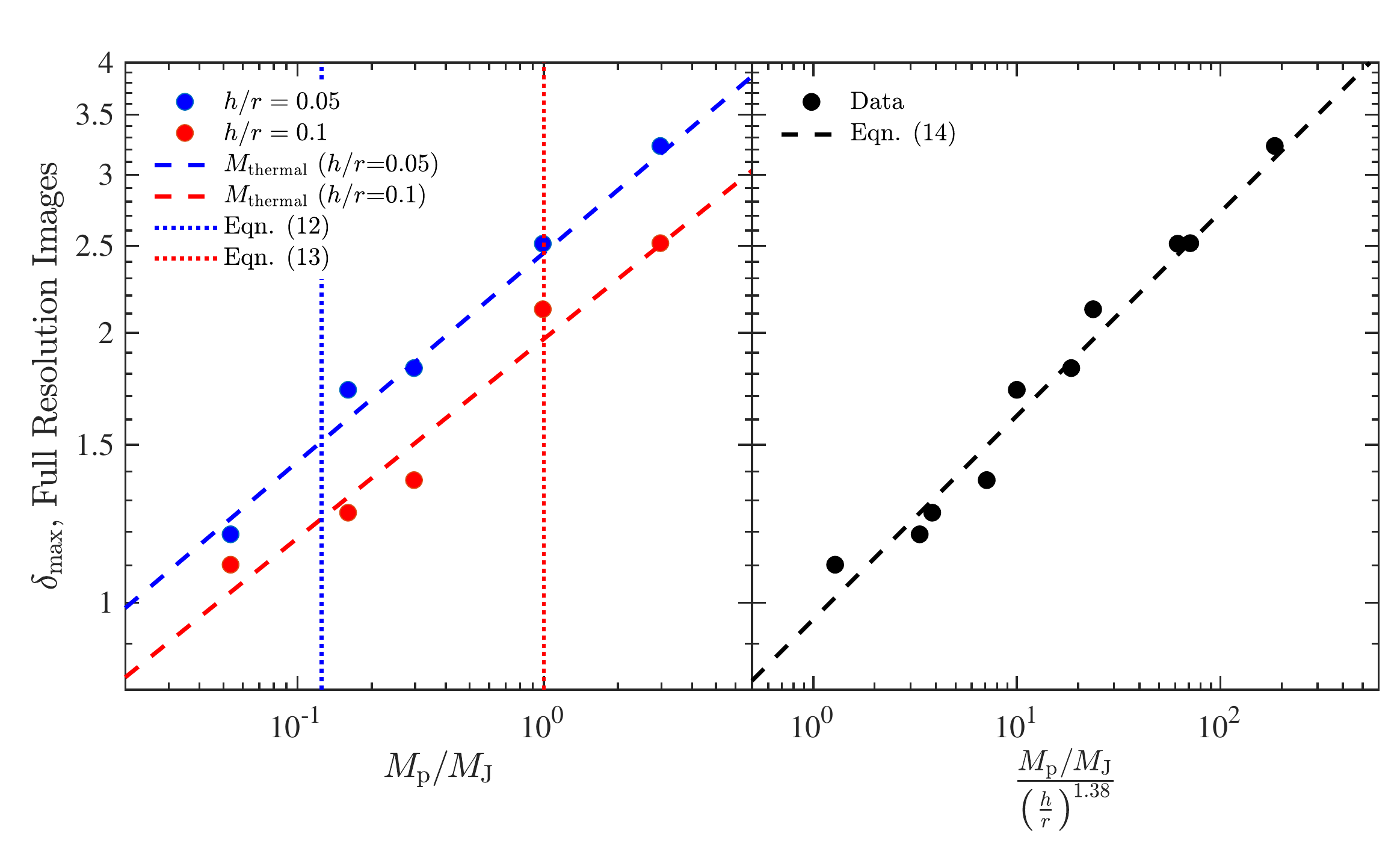}
\end{center}
\figcaption{Arm contrast $\delta_{\rm max}$ (the global maximum of Equation~\ref{eq:contrast_r} for the inner arms as a function of planet mass for the full resolution images. {\bf Left panel} shows data points for the $h/r=0.05$ and $h/r=0.1$ series separately, while the  {\bf right panel} shows the two series together with a horizontal axis of $\mplanet/(h/r)^{1.38}$. Fitted results (Equations~\ref{eq:contrast_h5}-\ref{eq:contrast}) are over plotted as dashed lines. $\mt$ for both $h/r$ are marked by the vertical dotted lines on the left panel.
\label{fig:contrast_raw}}
\end{figure}

\begin{figure}
\begin{center}
\includegraphics[trim=0 0 0 0, clip,width=0.5\textwidth,angle=0]{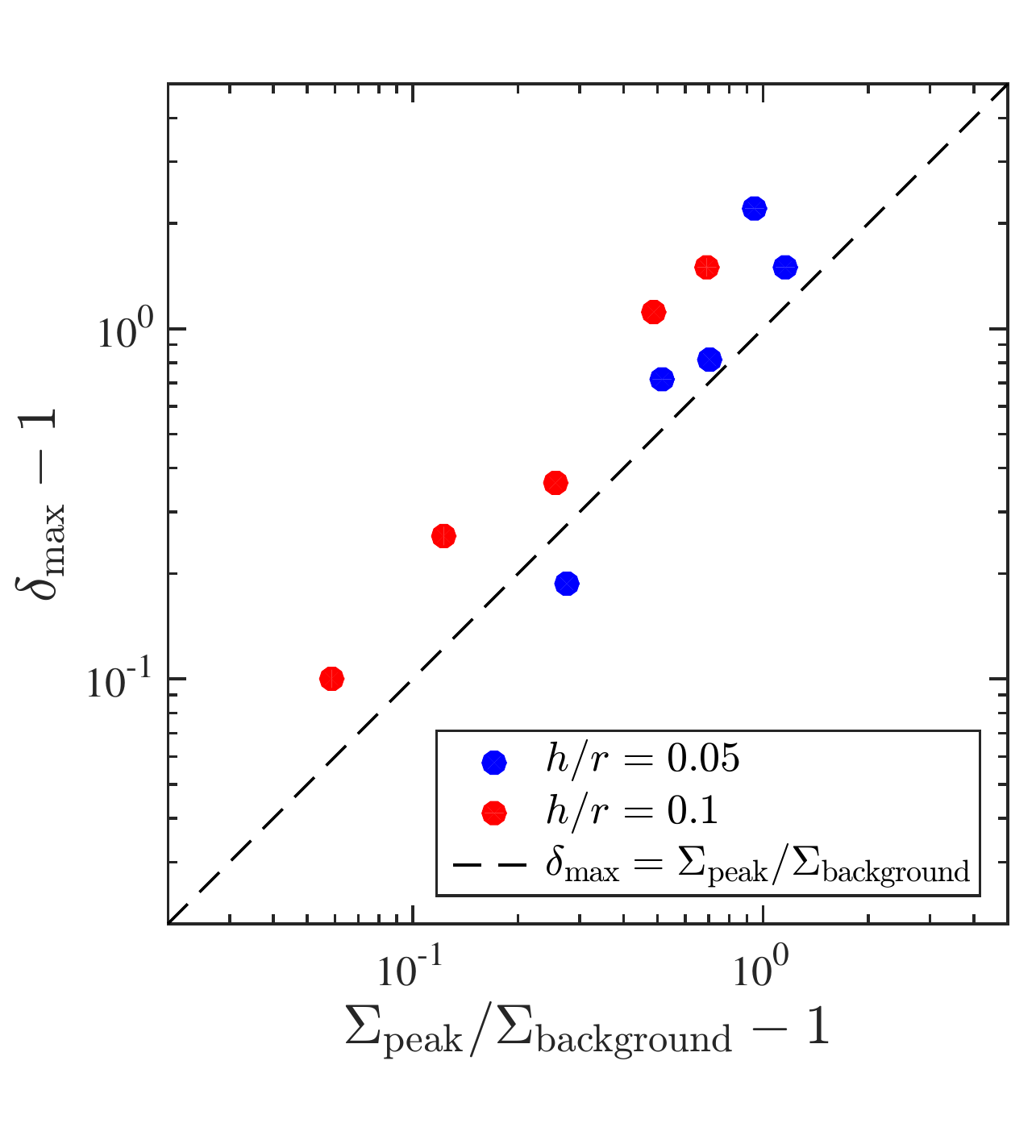}
\end{center}
\figcaption{Comparison between arm contrast in surface density with arm contrast in scattered light. $\Sigma_{\rm peak}/\Sigma_{\rm background}$ is measured at the same location as $\delta_{\rm max}$. In general, the structures in surface density are closely matched to the ones observed in the scattered light images.
\label{fig:contrast_sigma_image}}
\end{figure}

\begin{figure}
\begin{center}
\includegraphics[trim=0 0 0 0, clip,width=0.6\textwidth,angle=0]{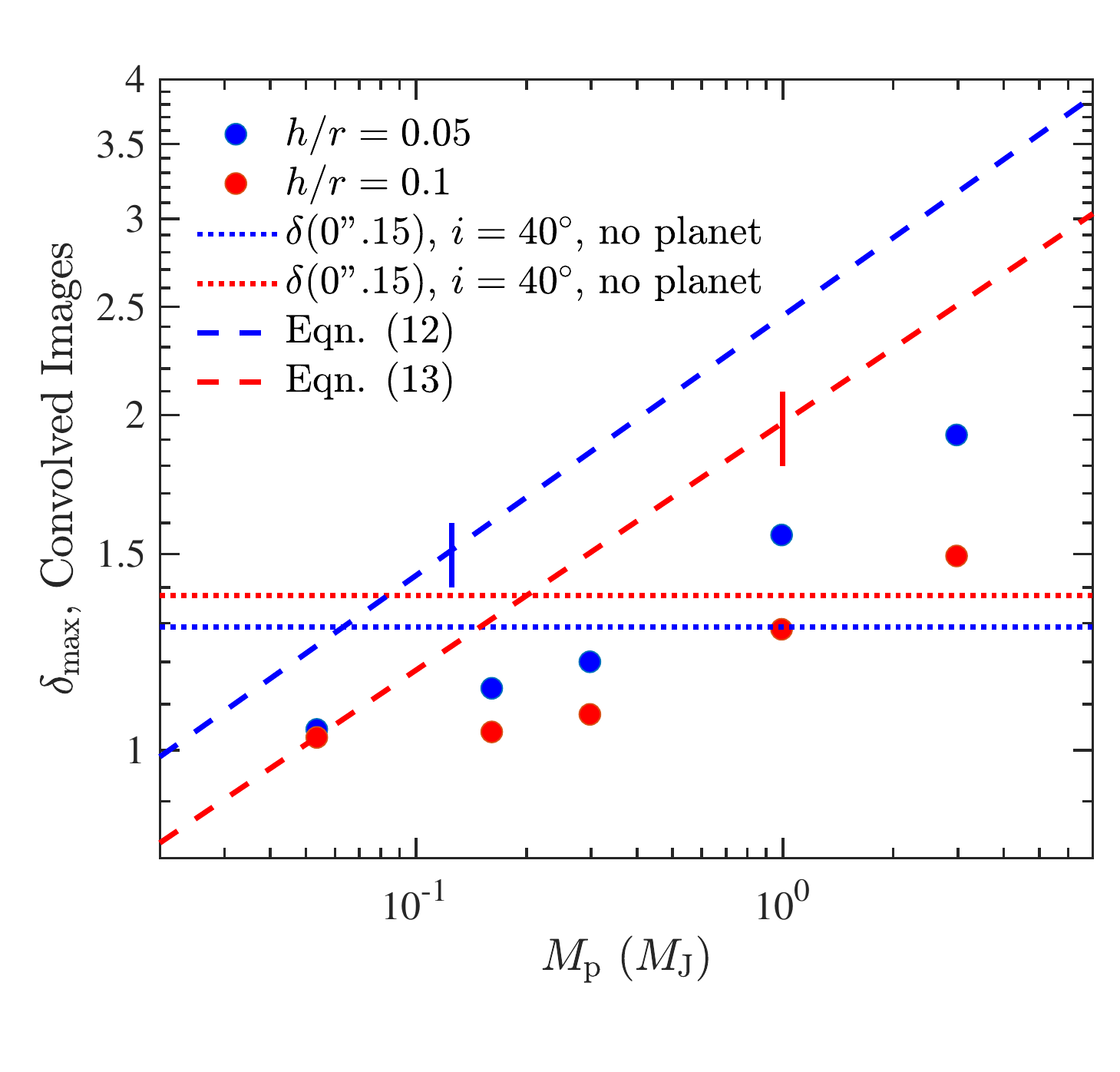}
\end{center}
\figcaption{Same as the left panel in Figure~\ref{fig:contrast_raw}, but for convolved images (each radial profile in Figure~\ref{fig:radial_max} contributes one data point). Horizontal dotted lines indicate the level of azimuthal variation at a projected radius $r=0\arcsec.15$ caused by an inclination $i=40^\circ$ in a disk with no planets under $h/r=0.05$ (blue) and $h/r=0.1$ (red). $\mt$ for both $h/r$ are marked as ticks on the dashed lines.
\label{fig:contrast_conv}}
\end{figure}

\begin{figure}
\begin{center}
\includegraphics[trim=0 0 0 0, clip,width=\textwidth,angle=0]{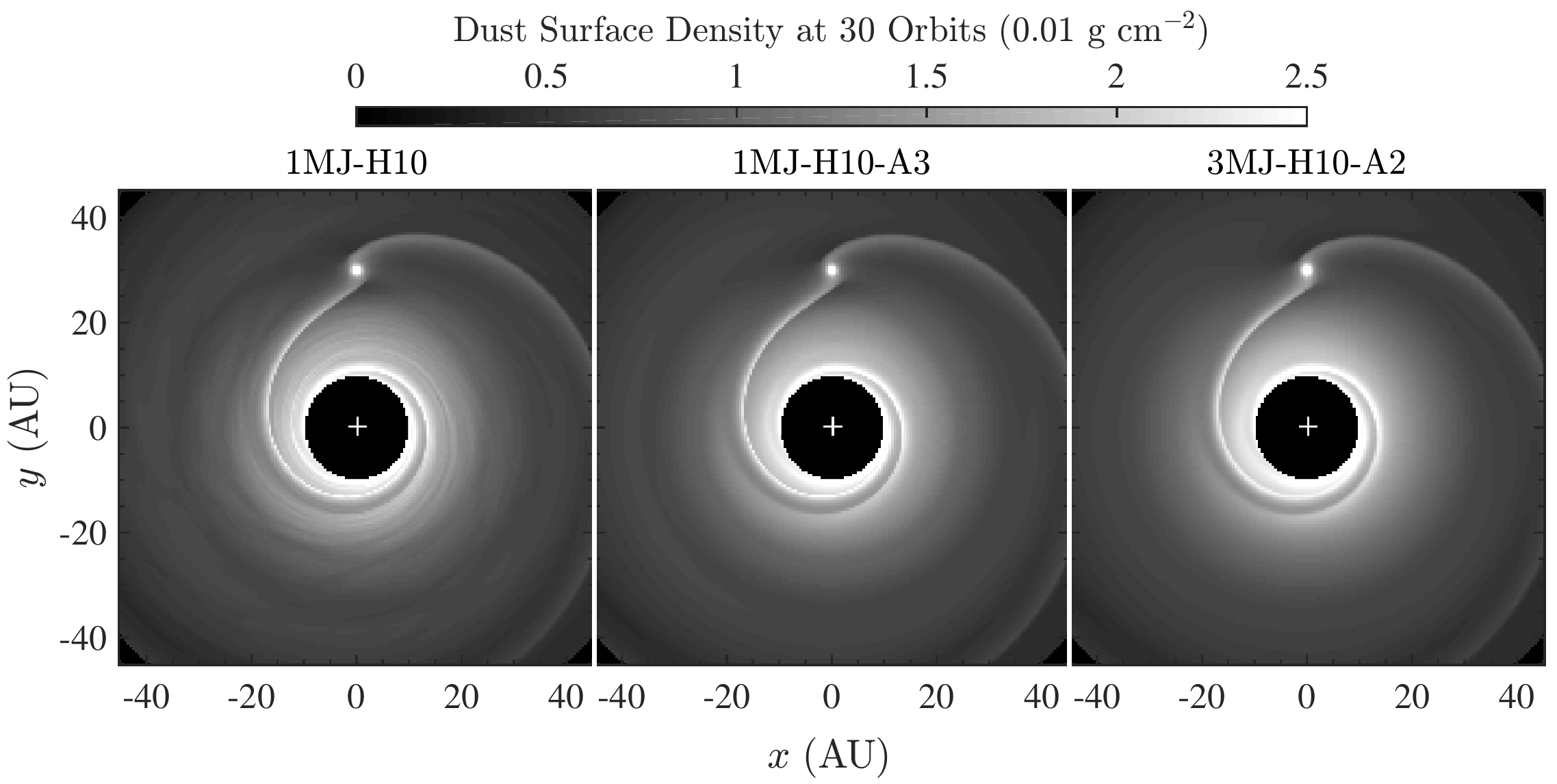}
\includegraphics[trim=0 0 0 0, clip,width=\textwidth,angle=0]{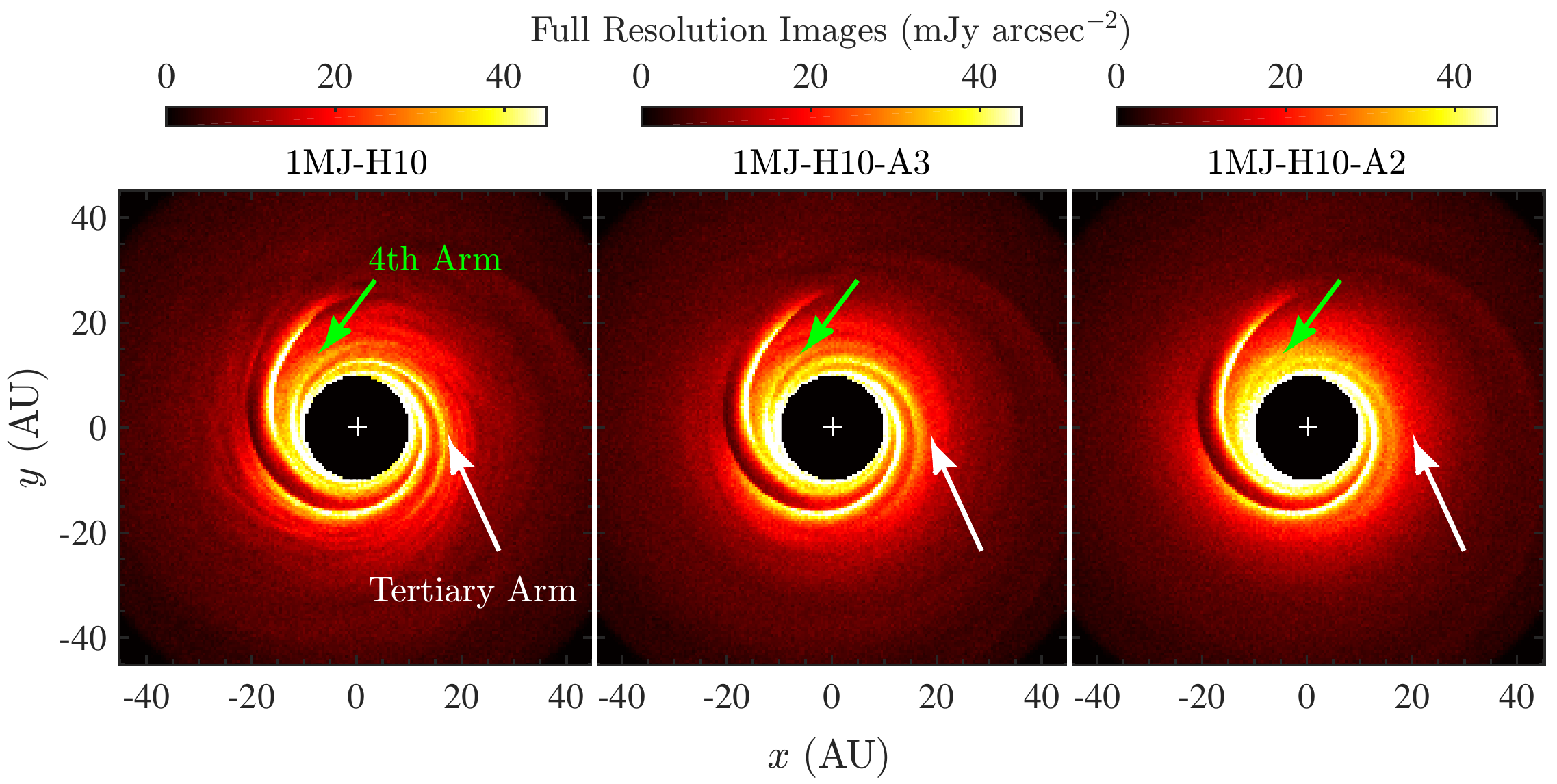}
\end{center}
\figcaption{Surface density ({\bf top row}) and full resolution $H$-band PI images ({\bf bottom row}) of 3 models with $h/r=0.1$, $\mplanet=1\mj$, and different viscosities: ({\bf from left to right}) $\alpha=0$, $\alpha=10^{-3}$, and $\alpha=10^{-2}$. The tertiary and the $4th$ arms are indicated in the bottom row. As viscosity increases, the locations, shape, and contrasts of the primary and secondary arms remain roughly unchanged, while the additional arms are weakened. See Section~\ref{sec:viscosity} for details.
\label{fig:viscosity}}
\end{figure}

\begin{figure}
\begin{center}
\includegraphics[trim=0 0 0 0, clip,width=0.49\textwidth,angle=0]{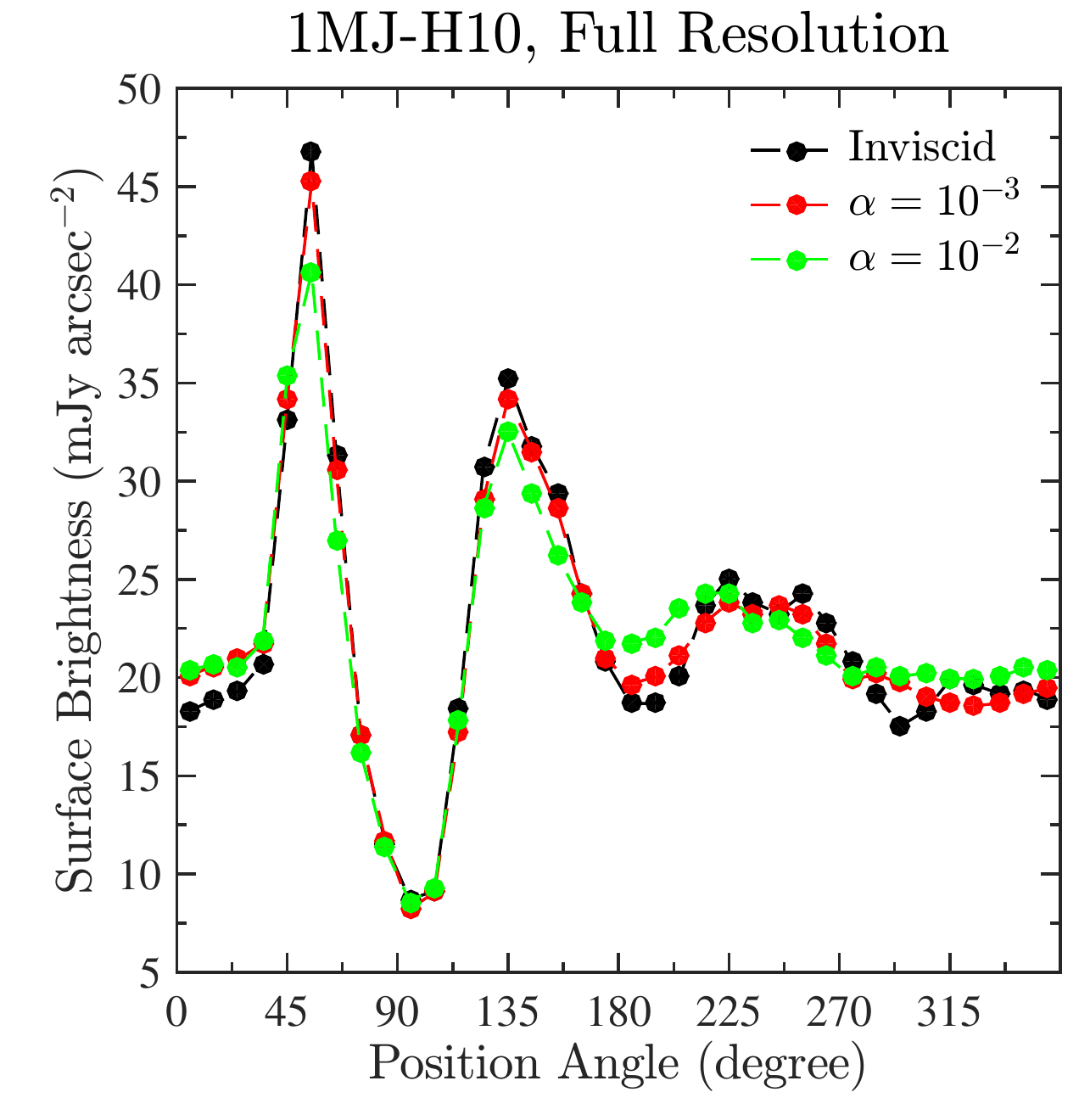}
\includegraphics[trim=0 0 0 0, clip,width=0.49\textwidth,angle=0]{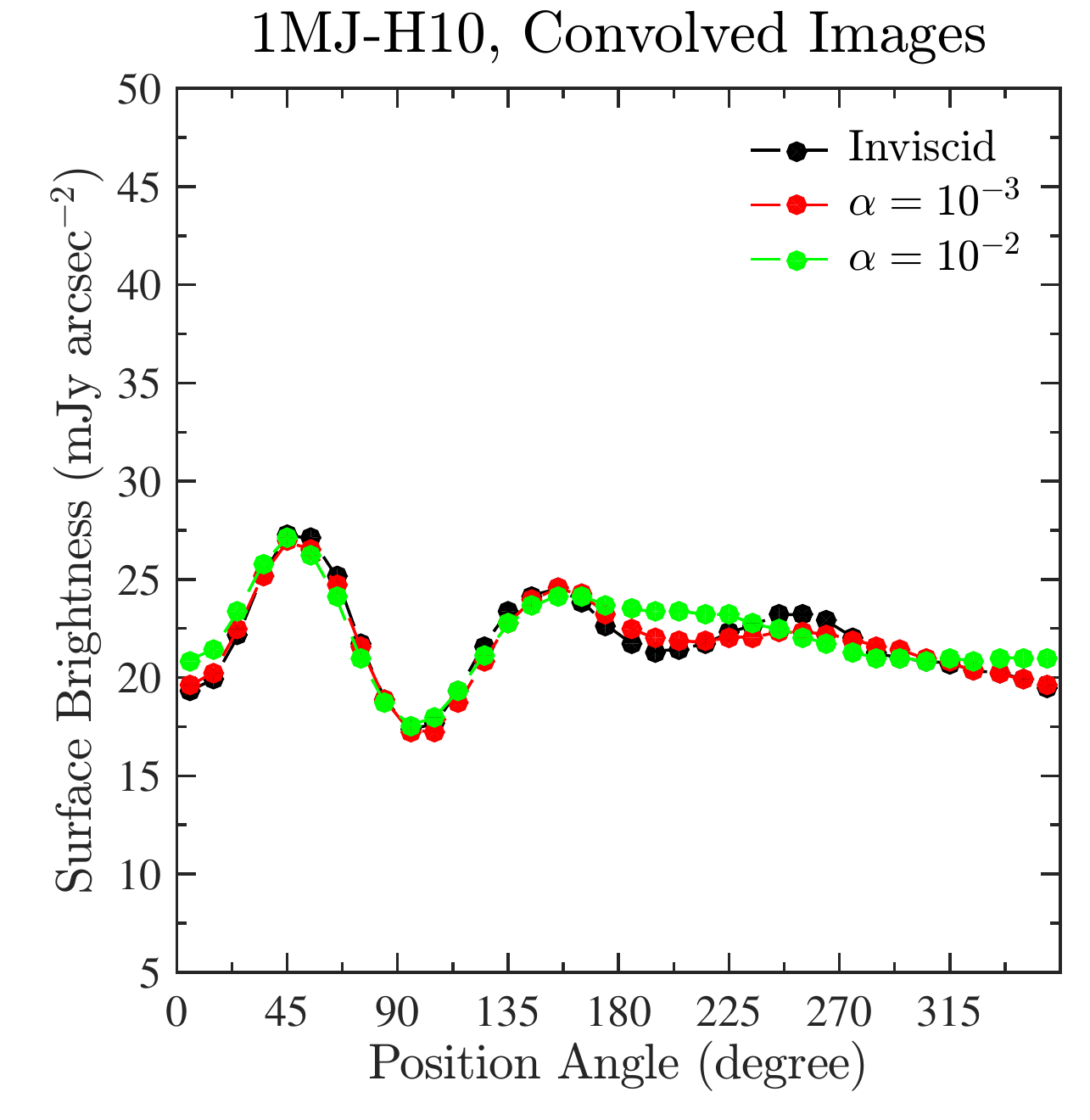}
\end{center}
\figcaption{The azimuthal arm profiles at $r=0\arcsec.13$ for three models (Left: full resolution images; Right: convolved images) with $\mplanet=1\mj$, $h/r=0.1$, and different viscosities. In the full resolution images, the peaks of the primary and secondary arms drop by $\lesssim10\%$ as $\alpha$ increases from 0 to $10^{-2}$, while the two remain essentially unchanged in the convolved images. The contrasts of the additional arms drop as viscosity increases. The locations of the arms, indicated by the position angles of the peaks, are essentially unchanged by viscosity. See Section~\ref{sec:viscosity} for details.
\label{fig:viscosity_azimuthal}}
\end{figure}

\begin{figure}
\begin{center}
\includegraphics[trim=0 0 0 0, clip,width=0.49\textwidth,angle=0]{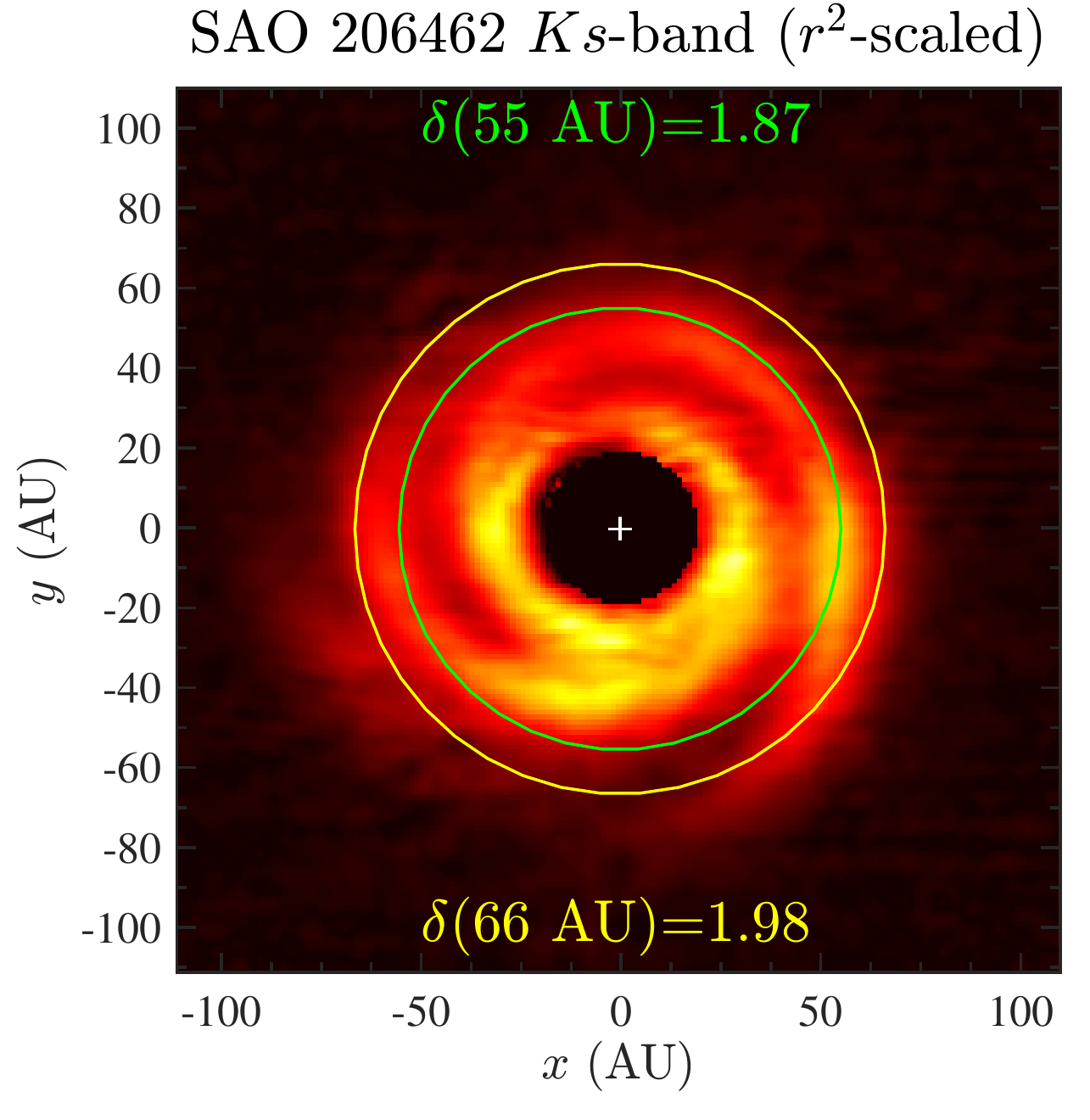}
\includegraphics[trim=0 0 0 0, clip,width=0.49\textwidth,angle=0]{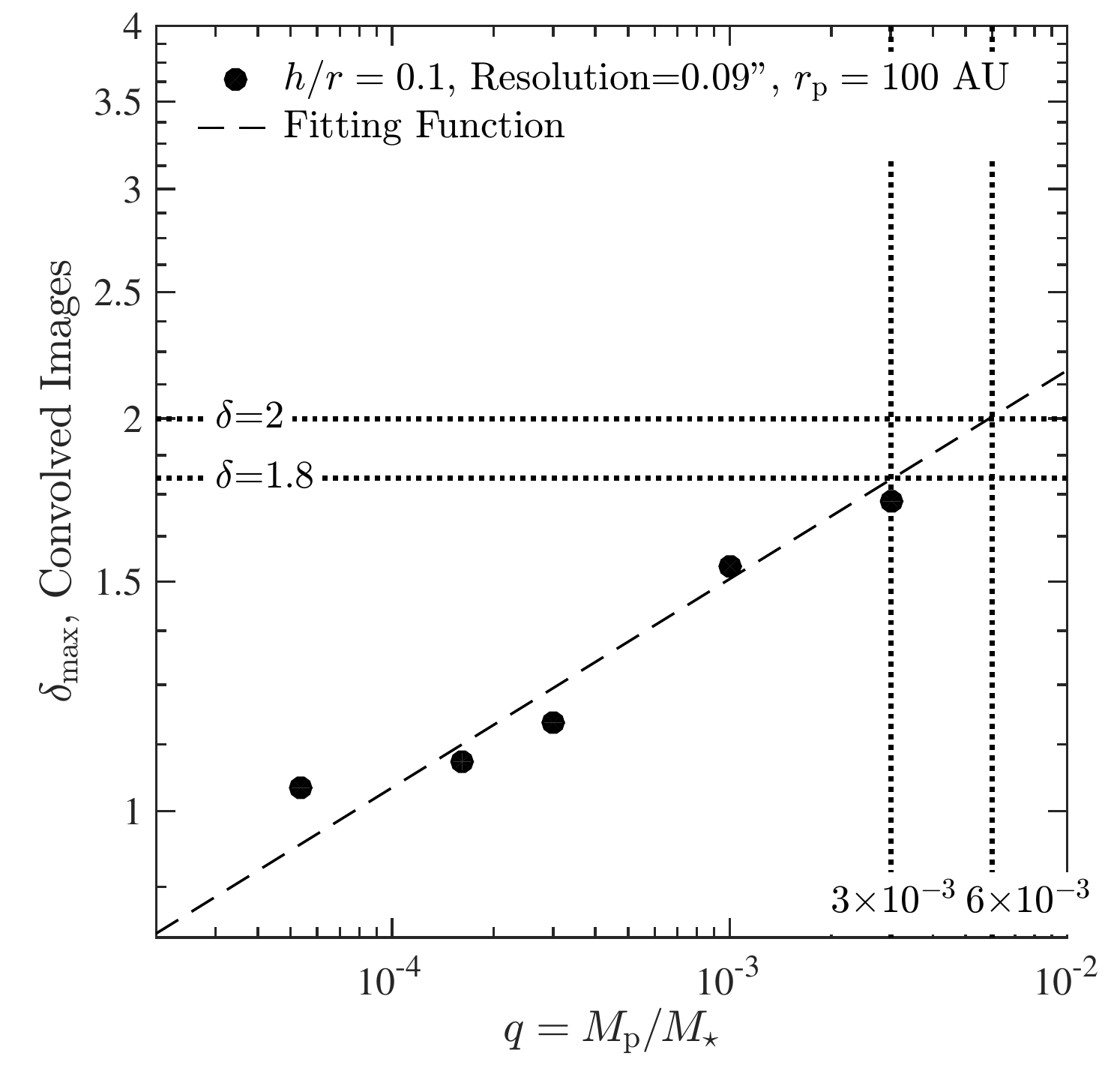}
\end{center}
\figcaption{{\bf Left:} VLT/NACO $Ks$-band image of SAO 206462 as presented in \citet{garufi13}. The arm contrast at 55 and 66 AU (indicated by the circle with the same color) are labeled in the image. {\bf Right:} $\delta_{\rm max}$ for our $h/r=0.1$ models after rescaling the planet to be at 100 AU and convolving to achieve an angular resolution of $0\arcsec.09$ to match the observation. The dashed-line is a simple power law fitting function of the data points: $y=4.6 x^{0.16}$. The two horizontal dotted lines bracket the range of $\delta(r)$ at the peak in SAO 206462, while the two vertical horizontal dotted lines bracket the range of possible planet mass of SAO 206462 b based on the arm contrast. See Section~\ref{sec:sao} for details.
\label{fig:sao}}
\end{figure}

\begin{figure}
\begin{center}
\includegraphics[trim=0 0 0 0, clip,width=0.9\textwidth,angle=0]{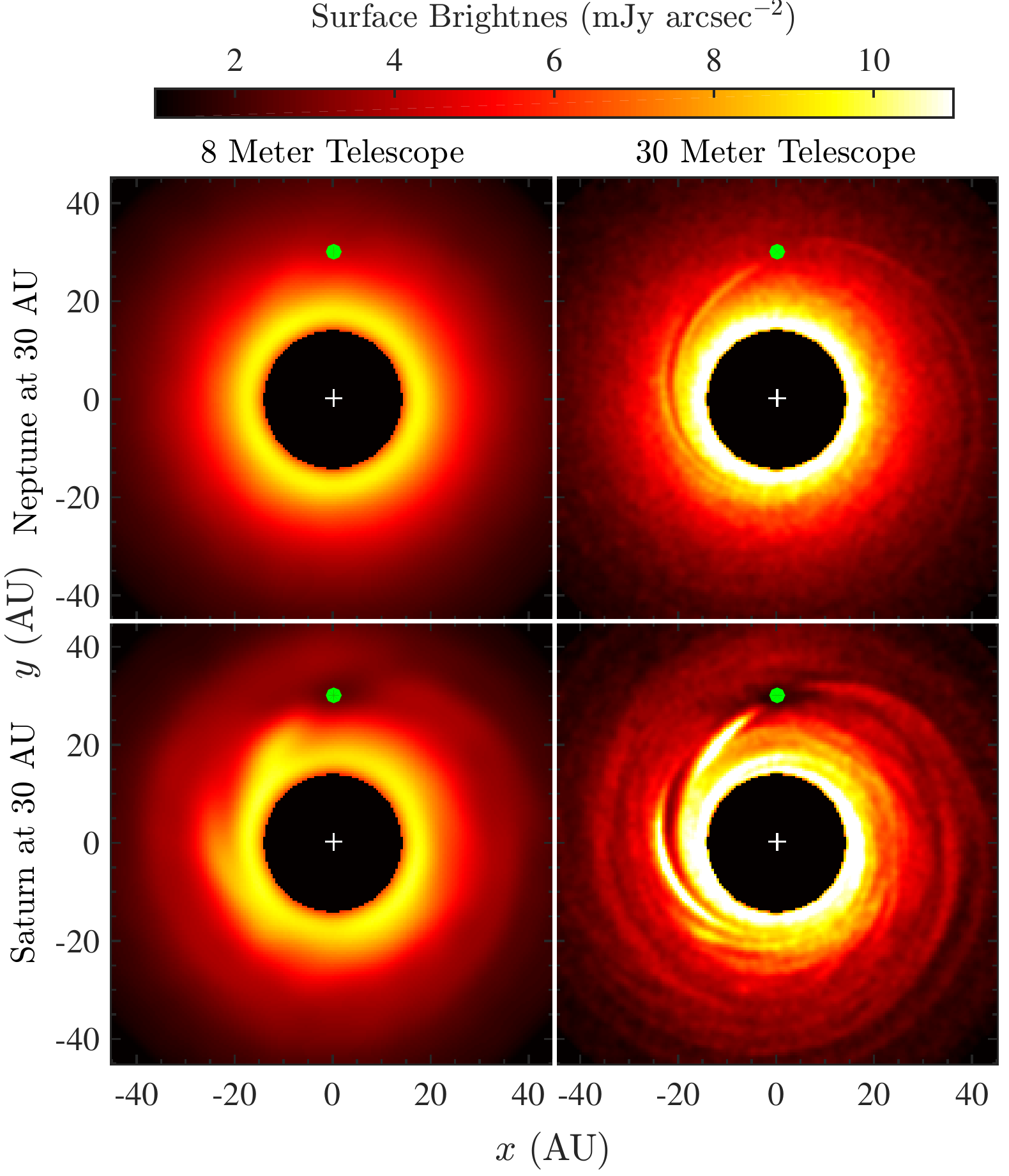}
\end{center}
\figcaption{Polarized intensity images of arms driven a Neptune ({\bf top row}) and a Saturn ({\bf bottom row}) mass planet at 30 AU (position marked by the green dot) in the $h/r=0.05$ series, convolved to reach the $H$-band angular resolution of an 8-m telescope ($0\arcsec.04$; {\bf left column}) and a 30-m telescope ($0\arcsec.01$; {\bf right column}). The inner $0\arcsec.1$ is masked out to emphasize the arms. With a 30-m telescope, the arms driven by a Neptune-mass planet is clearly visible; with an 8-m telescope, the arms driven by a Neptune-mass planet is not visible, while the arms driven by a Saturn-mass planet is only marginally recognizable. 
\label{fig:image_tmt}}
\end{figure}

\end{document}